\documentclass[fleqn,usenatbib]{mnras}

\usepackage[T1]{fontenc}
\usepackage{ae,aecompl}

\usepackage{longtable,lscape}
\usepackage{multicol}
\usepackage{afterpage}
\usepackage{natbib}
\usepackage{txfonts}
\usepackage{hyperref}
\usepackage[labelformat=simple,labelfont=bf,labelsep=period]{caption}
\usepackage{booktabs}

\def\m2s2{\,m$^{2}$\,s$^{-2}$} %m2.s -2
\usepackage{graphicx}	% Including figure files
\usepackage{amsmath}	% Advanced maths commands
\usepackage{amssymb}	% Extra maths symbols
\usepackage{multirow}
\usepackage{afterpage}
\usepackage{pdflscape}

\title[Direct Imaging Target Selection with COPAINS]{A New Method for Target Selection in Direct Imaging Programs with COPAINS}

\author[C. Fontanive et al.]{
C. Fontanive$^{1,2}$\thanks{E-mail: \href{mailto:clemence.fontanive@csh.unibe.ch}{clemence.fontanive@csh.unibe.ch}},
K. Mu{\v z}i{\'c}$^{3}$,
M. Bonavita$^{2}$
and B. Biller$^{2}$\\
\\
$^{1}$ Center for Space and Habitability, University of Bern, Gesellschaftsstrasse 6, 3012 Bern, Switzerland\\
$^{2}$ SUPA, Institute for Astronomy, University of Edinburgh, Blackford Hill, Edinburgh EH9 3HJ, UK\\
$^{3}$ CENTRA, Faculdade de Ci\^{e}ncias, Universidade de Lisboa, Ed. C8, Campo Grande, P-1749-016 Lisboa, Portugal
}

\date{Accepted 2019 September 7. Received 2019 September 3; in original form 2019 July 25.}

\pubyear{2019}

\begin{document}

\label{firstpage}
\pagerange{\pageref{firstpage}--\pageref{lastpage}}
\maketitle

% Abstract of the paper
\begin{abstract}
We present COPAINS (Code for Orbital Parametrisation of Astrometrically Inferred New Systems), an innovative tool developed to identify previously undiscovered companions detectable via direct imaging, based on changes in stellar proper motions across multiple astrometric catalogues. This powerful procedure allows for dynamical predictions of the possible masses and separations of unseen companions compatible with observed astrometric trends, marginalised over unknown orbital elements. Validating our approach using well-constrained systems, we found that our tool provides a good indication of the region of the parameter space where undetected secondaries may be located. Comparing the output of the code to the measured or expected sensitivity from various imaging instruments, this in turn enables us to robustly select the most promising targets for direct imaging campaigns searching for low-mass companions. Such an informed selection method promises to reduce the null detection rates from current programs and will significantly increase the current census of wide brown dwarfs and planetary companions to stars, which remain extremely rare in the surveys conducted so far.
\end{abstract}

% Select between one and six entries from the list of approved keywords.
% Don't make up new ones.

%%%%%%%%%%%%%%%%% BODY OF PAPER %%%%%%%%%%%%%%%%%%

\section{Introduction}
\label{intro}

The majority of known exoplanets and brown dwarf companions to stars were discovered using indirect detection techniques, such as the radial velocity (RV) and transit methods \citep{Charbonneau2007,Marcy2008,Mayor2008}. These approaches require observations covering multiple full orbital periods to confirm candidates, and are thus not suitable to identify wide-orbit companions, on orbital separations larger than a few AU. Direct imaging, on the other hand, is sensitive to companions $>1$ M$_\mathrm{Jup}$, at separations of tens to thousands of AU (see \citealp{Bowler2016} for a review).

A dozen of directly-imaged exoplanets with masses below the deuterium-burning limit have been confirmed around main sequence stars, and roughly twice as many companions with masses in the brown dwarf mass range are known. Despite these discoveries, previous surveys have generally yielded null detections or had high false positive rates, as a result of the low occurrence rates of wide brown dwarfs and giant planets, combined with the current limitations of high-contrast imaging instruments \citep{Biller2007,Lafreniere2007,Metchev2009,Nielsen2010,Vigan2012,Biller2013,Rameau2013,Galicher2016,Vigan2017}.
With little information available about the demographics of these populations, target selection is a particularly challenging process. Direct imaging programs typically target young stars ($<100$ Myr) to assure that the planets to be imaged are still bright, thus increasing the chances of detection, but selected stars generally have no a priori evidence for companions. Formation theories can provide indications of stellar properties that might be correlated or more favourable to planet formation (e.g. stellar mass, metallicity; see \citealp{Alibert2011,Mordasini2012}) and samples can be gathered based on such stellar characteristics.

Alternatively, a more efficient approach would consist in identifying independent physical signs of the presence of a hidden companion to select promising targets in imaging campaigns. For example, in the TRENDS survey, \citet{Crepp2012} searched for direct imaging companions to stars exhibiting long-term radial velocity accelerations (see also e.g. \citealp{Kasper2007,Janson2009,Rodigas2016}). However, as RV is unsuitable for young, active stars, these campaigns focused on older stars and the companions imaged are mostly in the stellar and substellar regimes. Precision astrometry can also be used to investigate nonlinear changes in stellar positions induced by an unseen companion. A significant discrepancy in proper motion measurements between different catalogues is a good indication of the presence of a perturbing body. Such proper motion accelerations may be used to select potential binaries (e.g. \citealp{Makarov2005,Tokovinin2013}), although so far, this method has mostly been restricted to stellar binaries due to the limited precision of astrometric measurements. 

In this paper, we present a new tool developed to address the target selection problems encountered in direct imaging studies: COPAINS (Code for Orbital Parametrisation of Astrometrically Inferred New Systems). By exploiting the synergy between direct imaging and astrometry, our innovative approach uses proper motion differences to provide a robust, informed selection method for favourable systems to observe in direct imaging searches based on changes in stellar proper motions. We introduce astrometric binaries and the fundamental principles of our approach in Section \ref{dmu_binaries}. In Section \ref{success_rate}, we investigate the success rate of such a target selection procedure. We detail the methodology of our tool in Section \ref{methodology} in which we present our target selection method in the context of direct imaging searches. In Section \ref{limitations}, we provide a detailed assessment of the efficiency of this new selection method. We present examples of applications of our approach to known directly-imaged targets in Section \ref{application}. Our results on the performance of our new tool are summarised in Section \ref{conclusion}.

%\begin{figure*}
%    \centering
%    \includegraphics[width=0.6\textwidth]{COPAINS_logo_wtext.pdf}
%    \caption{Logo of COPAINS.}
%    \label{f:logo}
%\end{figure*}

\section{$\Delta\mu$ astrometric binaries}
\label{dmu_binaries}

\begin{figure}
    \centering
    \includegraphics[width=0.42\textwidth]{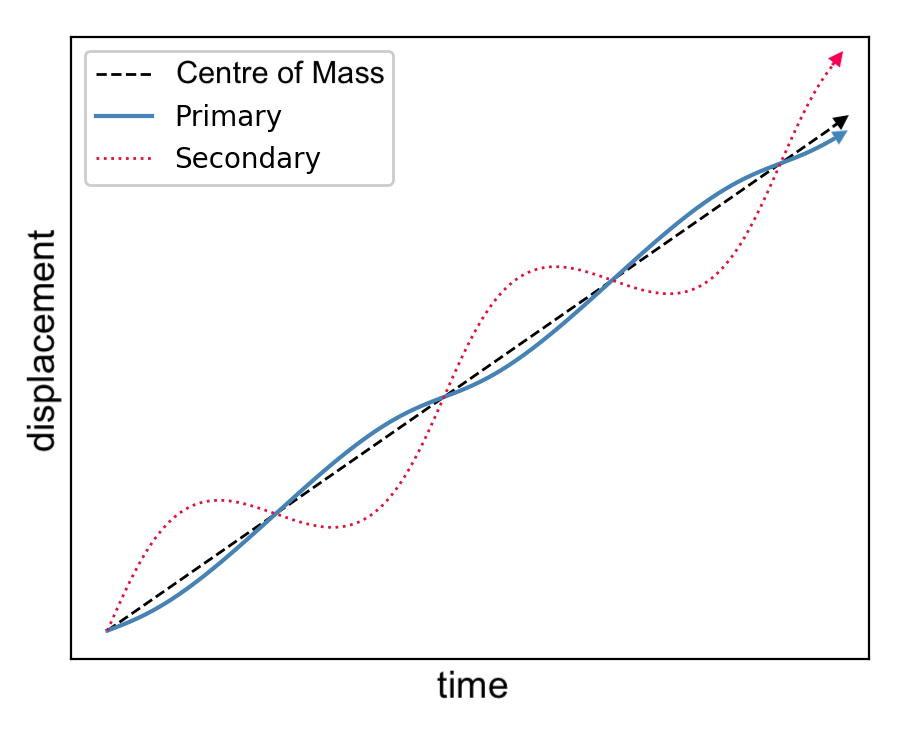}
    \caption{Example of the astrometric wobble of binary components around the centre of mass of the system. Measurements of the displacement of the primary star (solid blue line) relative to the centre of mass (dashed black line) can be used to infer the existence of the hidden secondary companion (dotted red line).}
    \label{f:astrometric_wobble}
\end{figure}

The wobble of a star in its orbit induced by the gravitational pull of a companion is one of the most efficient ways to detect the presence of that companion. When this wobble is (partly) along the line of sight, it can be observed through a periodic Doppler effect in the radial velocity of the host star. Hundreds of radial velocity planets have been identified via this technique, mostly on short orbital separations, where the method is most successful \citep{Charbonneau2007}. When the movement of the host star is in the plane of the sky, it can be visible as a change in the star's apparent position. This effect is illustrated in Figure \ref{f:astrometric_wobble}, which shows the astrometric displacement of the two components of a binary around the centre of mass of the system. A nonlinear apparent motion for a star may thus serve as evidence for the existence of a gravitationally bound, unseen companion.

The long orbital periods of most directly-imaged companions, together with the large jitter induced by the strong activity of young stars, make high precision radial velocity measurements extremely challenging for these targets. In contrast, the detectability of astrometric signatures increases with separation, while youth and stellar activity are not an issue for positional measurements. This makes astrometry highly compatible with direct imaging in terms of the optimal parameter space probed. Changes in stellar positions have successfully been used to identify stellar binaries \citep{Makarov2005,Tokovinin2012}, although up to now, astrometric determinations have been too inaccurate to detect lower-mass companions inside the substellar regime.

The difference $\Delta\mu$ between the instantaneous velocity (tangential vector to any point on the blue curve in Figure \ref{f:astrometric_wobble}) of a star of mass $M_1$ and the true barycentric motion (dashed black line) is given by Equation \ref{eq:dmu} \citep{Makarov2005}:
\begin{equation}
    \Delta\mu \leq \frac{2\pi \, \varpi \, \mathcal{R}_0 M_2}{\sqrt{a \, M_{tot}}},
\label{eq:dmu}
\end{equation}
where $\varpi$ is the parallax of the system (in mas), $M_2$ the mass of the secondary (in M$_\odot$), $M_{tot} = M_1 + M_2$ the total mass of the system (in M$_\odot$), and $a$ the semi-major axis of the reduced mass system (in AU; equal to the sum of the semi-major axes of the individual components), providing a $\Delta\mu$ in mas yr$^{-1}$. $\mathcal{R}_0$ is the time-dependent orbital phase factor, given by the eccentricity $e$ and the eccentric anomaly $E$:
\begin{equation}
    \mathcal{R}_0 = \left( \frac{1 + e \cos{E}}{1 - e \cos{E}} \right)^{1/2}.
\label{eq:R0}
\end{equation}
The inequality in Equation \ref{eq:dmu} arises from projection effects in cases of inclined orbits relative to the plane of the sky. More massive components are required to produce the same projected astrometric deviations compared to the face-on case \citep{Makarov2005}. As a result, Equation \ref{eq:dmu} provides an upper limit on the observable offset in apparent velocity for systems in inclined orbital planes. For face-on orbits, the inequality becomes an exact equation.

Short-term proper motion measurements such as those provided by the \textit{Hipparcos} mission ($\sim$3.3-yr baseline; \citealp{ESA1997}) and the \textit{Gaia} DR2 catalogue ($\sim$1.8-yr time-span; \citealp{GaiaCollaboration2018}) can capture this reflex orbital motion caused by a companion on a sufficiently long period. On the other hand, long-term astrometric measurements like those from the Tycho-2 catalogue (almost a century timescale; \citealp{Hog2000}) and the Tycho-\textit{Gaia} Astrometric Solution (TGAS, $\sim$25-yr baseline; \citealp{Michalik2015}) subset of the \textit{Gaia} DR1 catalogue \citep{GaiaCollaboration2016} can provide values which are closer to the true centre-of-mass motion of the system. In this case, the difference in proper motion between a short and a long-term astrometric catalogue will be well approximated by the $\Delta\mu$ variable given in Equation \ref{eq:dmu}. A significant discrepancy in proper motion between measurements acquired over very different time baselines can therefore probe the acceleration of a star in its inertial frame of reference. This can in turn be used to highlight the presence of a hidden companion and be connected to the orbital properties of the system via Equation \ref{eq:dmu}. We refer to stars exhibiting significant proper motion offsets between various astrometric catalogues as $\Delta\mu$ stars.

Several thousands of such $\Delta\mu$ binaries have been identified by comparing the \textit{Hipparcos} and Tycho-2 catalogues (e.g. \citealp{Makarov2005, Frankowski2007}). Targeted searches have subsequently led to the imaging of the unseen companions, mostly stellar, compatible with the measured astrometric trends (see e.g. \citealp{Tokovinin2012,Tokovinin2013}).
Thanks to the excellent precision of the new ESA \textit{Gaia} satellite, the accuracy of astrometric measurements is finally sufficient to exploit stellar displacements due to the influence of lower-mass companions. This method can now be used to unveil much smaller trends, pointing towards the presence of companions in the brown dwarf and planetary mass regimes, well within the reach of current direct imaging technologies.

\section{$\Delta\mu$ Selection success rate}
\label{success_rate}

In this section, we quantify the success rate of a selection procedure based on this $\Delta\mu$ approach via two separate analyses. In Section \ref{Dmu_BF}, we investigate the binary fraction of $\Delta\mu$ stars relative to the overall stellar population, based on existing catalogues of stellar multiplicity and \textit{Gaia} DR2. In Section \ref{DI_selection}, we check whether our selection method would have selected targets with known directly-imaged substellar and planetary companions.

\subsection{Binary fraction of $\Delta\mu$ stars}
\label{Dmu_BF}

In order to estimate the multiplicity of $\Delta\mu$ stars compared to stars showing no significant changes in proper motion between long and short-term astrometric catalogues, we consider the Tycho-2 and \textit{Gaia} DR2 catalogues. The former catalogue is one of the largest archival data sets providing proper motion measurements acquired over timescales close to a century \citep{Hog2000}, while the highly precise astrometric data from \textit{Gaia} DR2 is based on only 22 months of observations \citep{Lindegren2018}. The combination of these two catalogues thus provides the best estimates of instantaneous and barycentre motions. $\Delta\mu$ measurements obtained from these catalogues are expected to be well approximated by Equation \ref{eq:dmu} for systems with a wide range of orbital periods. We note however that any proper motion discrepancies over time could in principle highlight the presence of a perturbing object, even if the observed changes in proper motion do not correspond to the $\Delta\mu$ in Equation \ref{eq:dmu}. As a result, any combination of astrometric catalogues can be used to identify stars with invisible companions. In particular, with the extremely high precision of \textit{Gaia} relative to previous astrometric missions, using the first two \textit{Gaia} Data Releases might provide the best chance at picking up on very small trends induced by very low-mass planetary companions, as discussed in Section \ref{DI_selection} below.

We obtained the entire cross-match of the Tycho-2 and \textit{Gaia} DR2 catalogues, and selected all targets with \textit{Gaia} DR2 parallaxes larger than 20 mas, to construct a volume-limited sample within 50 pc. We then removed all targets that lacked proper motion measurements in either catalogue, and excluded sources with excessive astrometric noise in \textit{Gaia}, adopting a 10\% relative precision criterion in parallax and proper motion as was done in \citet{GaiaCollaboration2018b}. This provided us with a sample of 8738 objects.

For each star in this sample, we then checked its multiplicity flag in the \textit{Hipparcos} main catalogue \citep{vanLeeuwen2007} and the Tycho-2 Double Star Catalogue (TDSC; \citealp{Fabricius2002}), in addition to its presence in the Catalog of Components of Double \& Multiple stars (CCDM; \citealp{Dommanget2000}) and Ninth Catalogue of Spectroscopic Binary Orbits (SB9; \citealp{SB9}). We also queried the position of the star in the \textit{Gaia} DR2 catalogue and searched for comoving sources within a angular radius corresponding to a projected separation of $10^4$ AU given the parallax of the star. We define comoving objects in \textit{Gaia} as sources with a fractional difference of $<20$\% in parallax and at least one of the proper motion coordinates, following the approach from \citet{Fontanive2019} to search for wide binary companions in \textit{Gaia} DR2. This ensures that binary components showing significant disparities in their kinematics due to their gravitational influence on one another (i.e. the $\Delta\mu$ stars of interest here) do get selected as binaries in this analysis. In a statistical search for wide binaries in TGAS based on parallaxes and proper motions, \citet{Andrews2017} found a contamination rate from random chance of alignment of $\lesssim5$\% for binaries with projected separations smaller than $4\times10^4$ AU. We therefore trust our binary selection criteria in \textit{Gaia} DR2 to select bonafide systems and consider that our resulting sample of \textit{Gaia} binaries is unlikely to be significantly polluted by unassociated random alignments.

We then calculated the total change in proper motion between Tycho-2 and \textit{Gaia} DR2 for each star, given by:
\begin{equation}
    \Delta\mu = \sqrt{\left(\mu_{\alpha*,\mathrm{T}} - \mu_{\alpha*,\mathrm{G}}\right)^2 + \left(\mu_{\delta,\mathrm{T}} - \mu_{\delta,\mathrm{G}}\right)^2 } \, ,
\label{eq:dmu_measured}
\end{equation}
where the subscripts T and G correspond to Tycho-2 and \textit{Gaia} DR2, respectively. The corresponding uncertainty $\sigma_{\Delta\mu}$ can then be calculated from the quadrature sum of the component uncertainties, following standard error propagation rules.

The fraction of multiple systems in the initial list of 8738 objects was found to be $34.6\pm0.6$ \%.
We found that 2407 stars had a $\Delta\mu$ with a significance larger than 3$\sigma$ (i.e. $\Delta\mu / \sigma_{\Delta\mu} > 3$). The subset of objects with significant $\Delta\mu$ trends had a multiplicity rate of $55.2\pm1.5$ \%, while the stars with no observed changes in proper motion only had a binary frequency of $26.5\pm0.6$ \%. With a final binary fraction more than twice as large for $\Delta\mu$ stars relative to the rest of the sample ($> 17$-$\sigma$ level), these results demonstrate the high efficiency of our $\Delta\mu$ approach to select targets for companion searches.

\begin{figure*}
    \centering
    \includegraphics[width=0.95\textwidth]{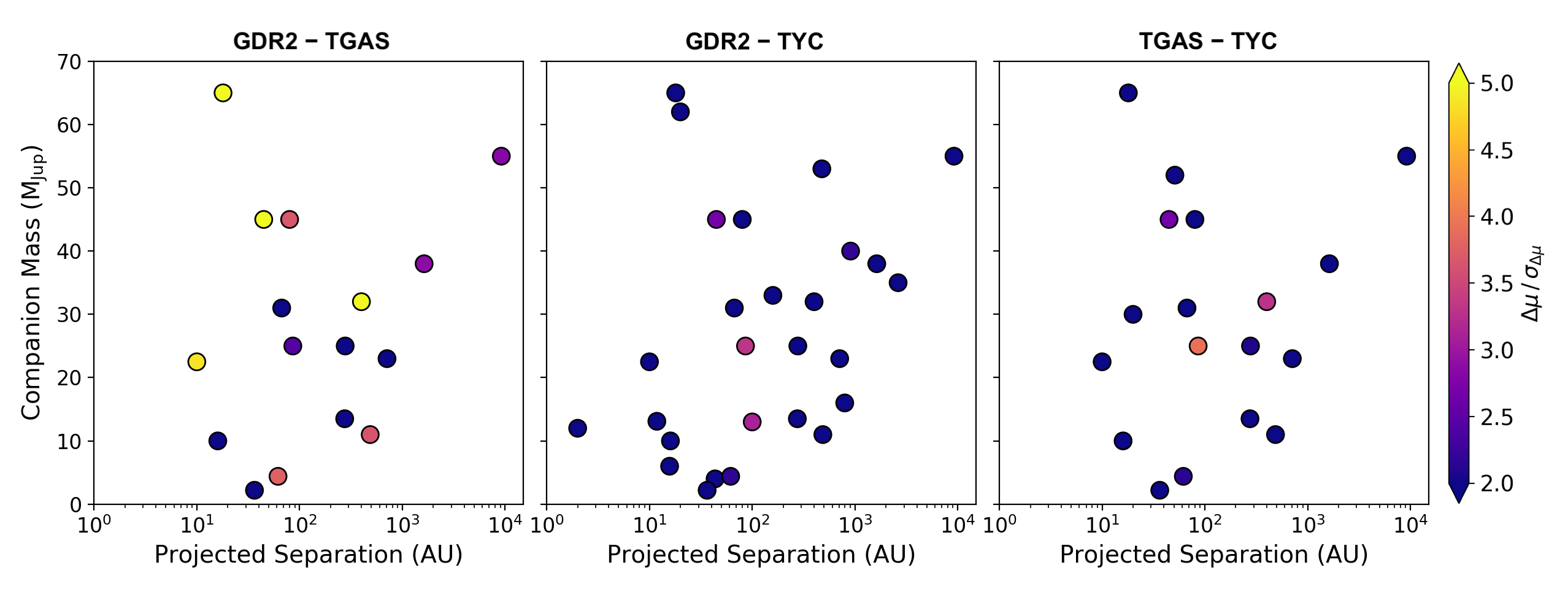}
    \caption{Measured changes in proper motion for 29 stars with directly-imaged brown dwarf or planetary companions, based on \textit{Gaia} DR2, TGAS and Tycho-2, showing the significance of measured $\Delta\mu$ values in the colour bar. About half of the available proper motion offsets between \textit{Gaia} DR2 and TGAS (left panel) are significant to more than 3$\sigma$ (from pink to yellow). In contrast, most $\Delta\mu$ values involving data from the Tycho-2 catalogue have a $<$ 2$\sigma$ significance, highlighted by the excess of blue symbols in the middle and right panels.}
    \label{f:DI_targets}
\end{figure*}

\subsection{Known targets with direct imaging companions}
\label{DI_selection}

The binary search performed in Section \ref{Dmu_BF} was mostly sensitive to stellar companions due to limitations in sensitivity of the catalogues considered. In order to establish whether the $\Delta\mu$ method for target selection is also valid for low-mass companions, inside the substellar regime, we compiled a list of host stars with directly-imaged brown dwarf and planetary companions, and checked if our approach would have selected these systems.

We gathered all objects from the Extrasolar Planet Encyclopaedia\footnote{\url{http://exoplanet.eu}} with a confirmed substellar or planetary companion discovered via direct imaging. We then queried the \textit{Gaia} DR2, TGAS and Tycho-2 proper motions and their associated errors for each target, and removed targets that lacked kinematic data in at least two of the three catalogues, providing a sample of 38 objects. We then checked if the stars in the remaining sample were flagged as binaries when performing the same analysis as in Section \ref{Dmu_BF}. We eliminated systems with an additional, tertiary component, as the gravitational influence from a more massive stellar companion would likely dominate the observed $\Delta\mu$ of the brown dwarf or planet host. This left us with a sample of 29 targets with no apparent higher-order companion, for which there was at least one, and up to three, $\Delta\mu$ measurements between the \textit{Gaia} DR2, TGAS and Tycho-2 catalogues.

Most of the programs in which these substellar companions were identified did not consider excursions in stellar astrometry in the target selection processes. We therefore consider that our sample is not biased towards or against the presence of detectable astrometric signatures. This means that the fraction of $\Delta\mu$ stars in this sample should be representative of the fraction of hosts to substellar companions that our approach can detect, given the precision of currently-available data.

Using Equation \ref{eq:dmu_measured} to compute proper motion disparities, we found that 9 stars in the final sample had at least one $\Delta\mu$ value significant to more than 3$\sigma$. Thus, about a third of the gathered sample would have been selected with this strategy. The most precise changes in proper motion came from the combination of the \textit{Gaia} DR2 and TGAS catalogues, as a result of the high accuracy achieved by \textit{Gaia} (typically $<0.1$ mas yr$^{-1}$) compared to previous missions. In particular, 16 targets in the sample had a \textit{Gaia} DR2-TGAS $\Delta\mu$ measurement, 7 of which had a significance $>$ 3$\sigma$, which represents the large majority of the significant proper motion offsets obtained in the full sample. In contrast, only 4 out of 45 changes in proper motion computed from Tycho-2 data were found to be above the 3-$\sigma$ threshold, due to the fact that the observed uncertainty $\sigma_{\Delta\mu}$ is higher for the Tycho-2 data, since the measurements are not as accurate. This is shown in Figure \ref{f:DI_targets}, which displays the significance of the proper motion measurements ($\Delta\mu / \sigma_{\Delta\mu}$) for each target with existing astrometric data in the two catalogues considered in each panel. We note that no obvious trend of detectability as a function of companion mass or separation is seen in Figure \ref{f:DI_targets}, although the inhomogeneity of the sample in stellar mass and distance means that the observed parameter space is not directly comparable between the various targets.

While these results are limited by the small size of the sample considered here, they suggest that the fraction of hosts to substellar companions that can be identified with a $\Delta\mu$ analysis increases significantly when using \textit{Gaia} data only. Given the small astrometric offsets induced on primaries by such low-mass companions (typically of only a few mas yr$^{-1}$ at most), it is not surprising that these signals are only rarely recovered when considering data like those provided by Tycho-2, with typical proper motion uncertainties of at least $\sim$1$-$2 mas yr$^{-1}$. The high precision of \textit{Gaia} DR1 and DR2 data, on the other hand, allows for the detection of much smaller trends, and can highlight the existence of substellar companions for $\sim$50\% of currently-known direct imaging systems with existing astrometric data in \textit{Gaia}.

This is extremely promising in anticipation of future \textit{Gaia} Data Releases, which will certainly improve these numbers and allow for robust detections of many more substellar companions to be observed with direct imaging facilities. Indeed, about a third of the stars gathered for this analysis were missing from the TGAS catalogue, hence the lower number of systems included in the left and right panels of Figure \ref{f:DI_targets}. With a higher completeness level and a superior astrometric precision, the upcoming \textit{Gaia} Data Releases will provide improved short-term proper motions, but also enable the calculation of new, highly-precise long-term proper motions by comparing new positional measurements to historical catalogues (see e.g. the \textit{Hipparcos}-\textit{Gaia} Catalog of Accelerations; \citealp{BrandtHGCA2018}). This will in turn allow us to highlight with a high significance $\Delta\mu$ trends induced by companions in the planetary regime, and for a larger number of stars than with a DR1-DR2 analysis.

\section{Methodology of the COPAINS tool}
\label{methodology}

\subsection{Computation of orbital properties}
\label{computation}

Based on Equation \ref{eq:dmu}, we can constrain the region of the parameter space in which a hidden companion responsible for an observed proper motion trend may be. For a given $\Delta\mu$ measurement, our COPAINS tool allows us to evaluate the possible companion mass and separation pairs compatible with the astrometric data, as a function of eccentricity.

\begin{figure*}
    \centering
    \begin{minipage}[b]{0.46\textwidth}
        \includegraphics[width=\textwidth]{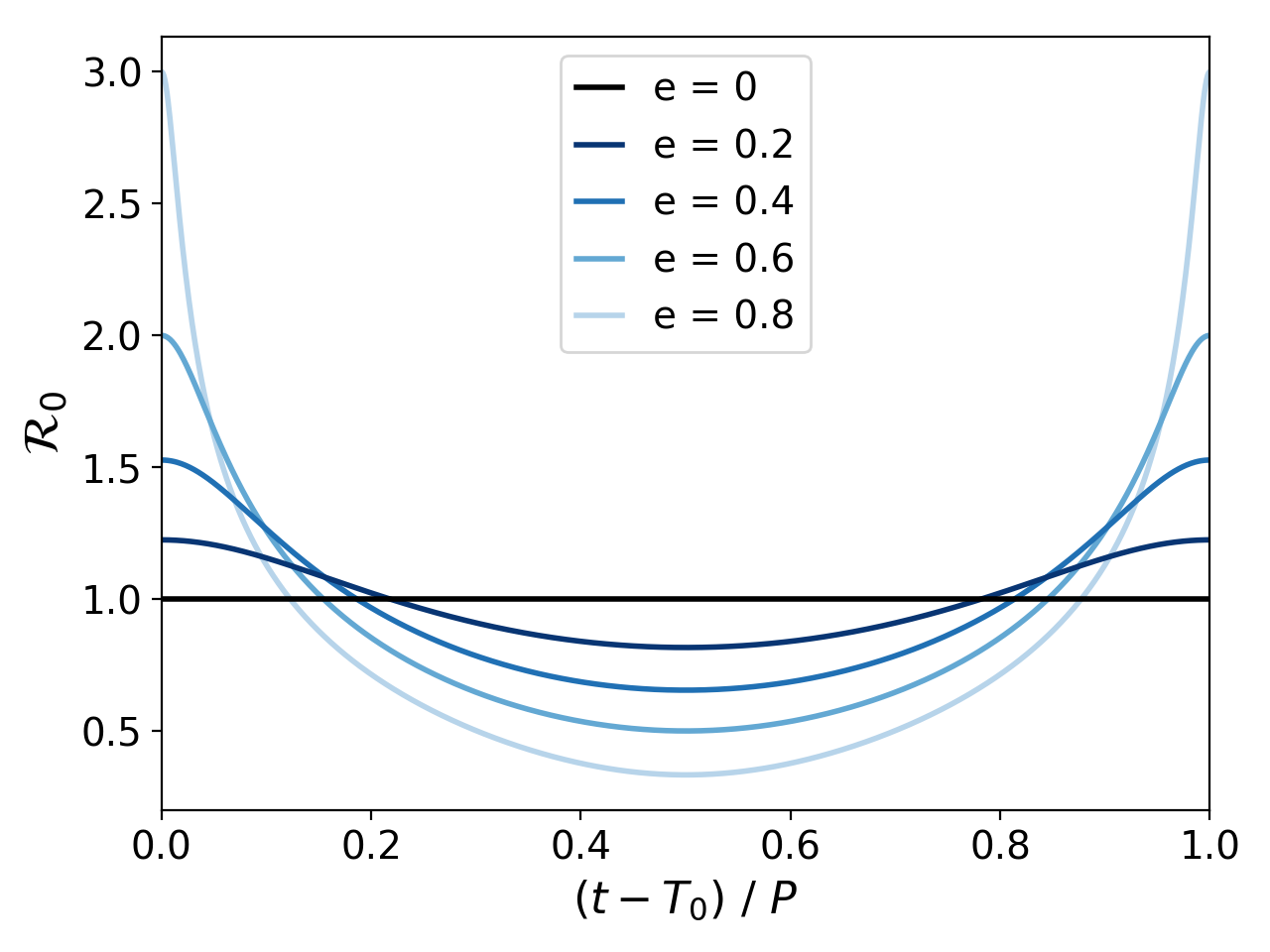}
    \end{minipage}
    \begin{minipage}[b]{0.46\textwidth}
        \includegraphics[width=\textwidth]{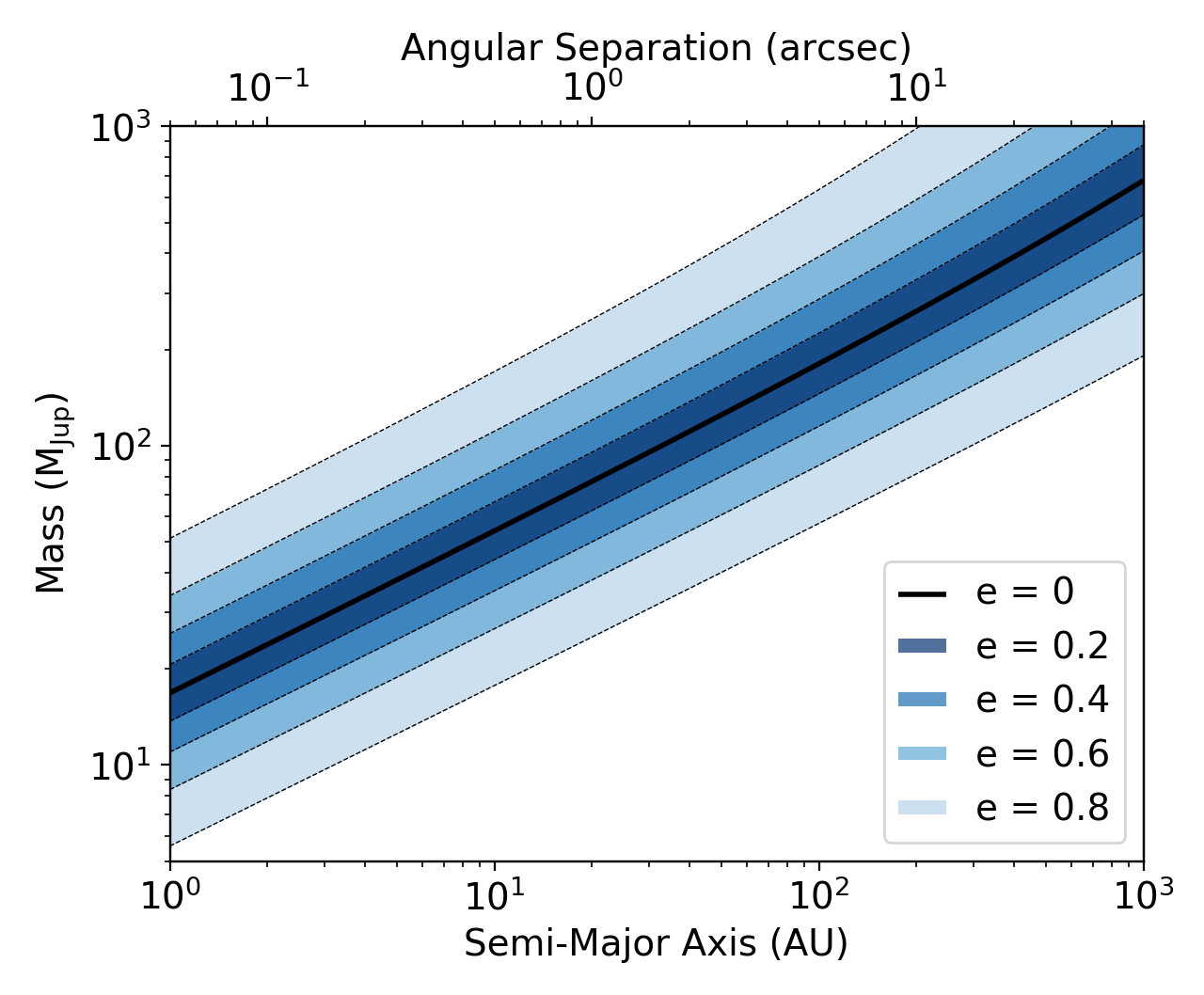}
    \end{minipage}
    \caption{\textbf{Left:} Orbital phase factor $\mathcal{R}_0$ as a function of the orbital phase $(t-T_0)/P$, for various fixed eccentricity values, calculated using Equation \ref{eq:R0} in the case of a face-on orbit. \textbf{Right:} Pairs of masses and semi-major axes for secondary companions compatible with the $\Delta\mu$ of an example star of mass $M_1=1$ M$_\odot$ and parallax $\varpi=50$ mas, with a proper motion offset of $\Delta\mu = 5\pm1$ mas yr$^{-1}$, for a face-on configuration. For each eccentricity, we plot the edges of the range of possible solutions, found by considering the minimum and maximum values allowed for $\mathcal{R}_0$ in the left panel, and using Equation \ref{eq:dmu} at those boundary $\mathcal{R}_0$ values.}
    \label{f:DMU_fixed_ecc}
\end{figure*}

We first investigate the behaviour of Equation \ref{eq:dmu} with eccentricity. For any fixed eccentricity value $e$, the orbital phase factor $\mathcal{R}_0$ (Equation \ref{eq:R0}) varies between a minimum and maximum value determined by the eccentric anomaly $E$. $E$ is itself a function of the time-dependent mean anomaly $\mathcal{M}$:
\begin{equation} \label{eq:anomalies}
    \mathcal{M} = 2\pi \, \frac{(t-T_0)}{P} = E - e\sin{E} ,
\end{equation}
where $T_0$ is the epoch of periastron passage and $P$ the orbital period of the system. $\mathcal{M}$ thus varies between 0 and 2$\pi$, and we can solve for $E$ over all possible $\mathcal{M}$ values and estimate the corresponding orbital phase factor $\mathcal{R}_0$ from Equation \ref{eq:R0}.
The left panel of Figure \ref{f:DMU_fixed_ecc} shows $\mathcal{R}_0$ as a function of the orbital phase $(t-T_0)/P$ for various eccentricity values, from which we can obtain the minimum and maximum values allowed for $\mathcal{R}_0$.

\begin{figure*}
    \centering
    \begin{minipage}[b]{0.445\textwidth}
        \includegraphics[width=\textwidth]{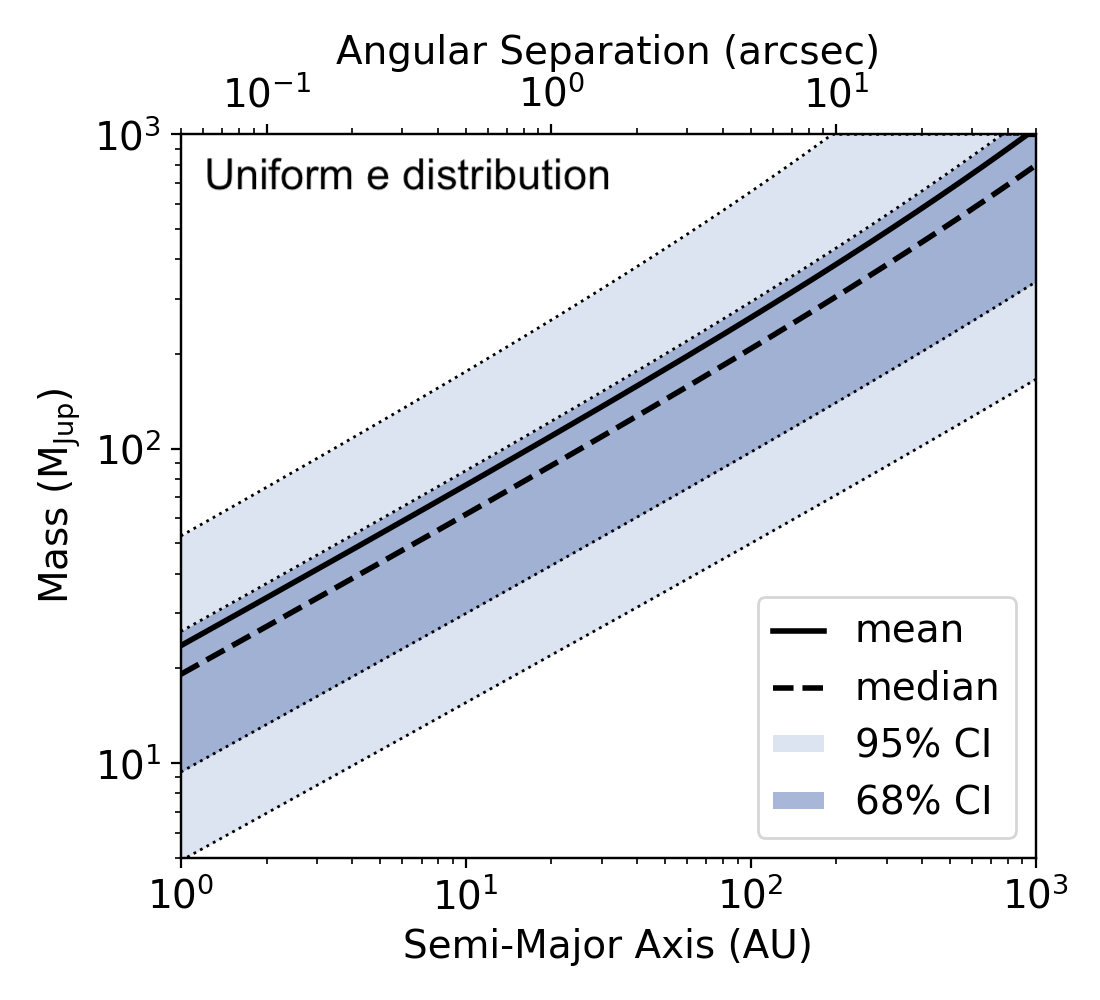}
    \end{minipage}  %\hfill
    \begin{minipage}[b]{0.445\textwidth}
        \includegraphics[width=\textwidth]{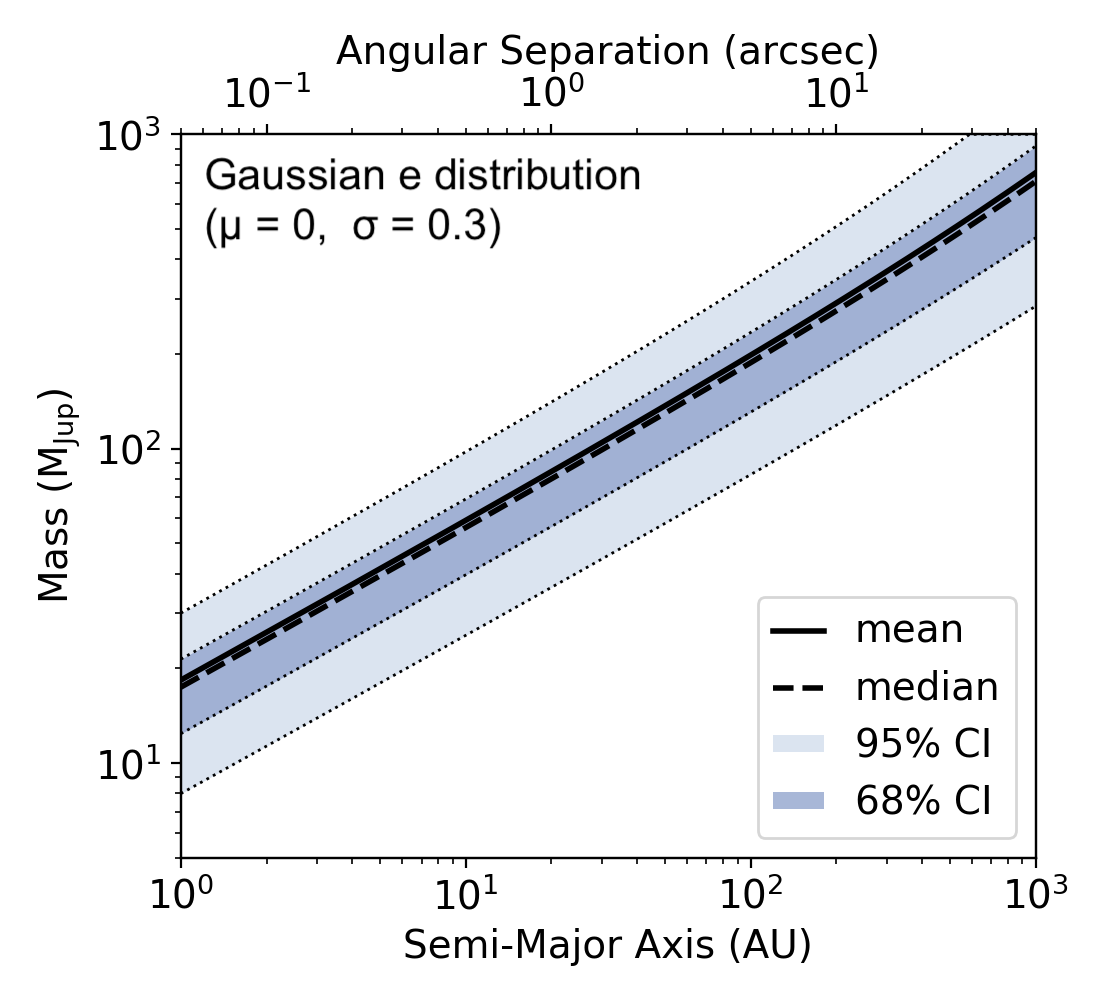}
    \end{minipage}
    \caption{Output of our COPAINS tool showing the 2-dimensional $M_2-a$ distribution for our example target of mass $M_1=1$ M$_\odot$ and parallax $\varpi=50$ mas, with an observed change in proper motion of $\Delta\mu=5\pm1$ mas yr$^{-1}$ for the primary, assuming a face-on orbit. In the left panel, we adopted a uniform distribution in eccentricity, while a Gaussian distribution of mean 0 and width 0.3 was used in the right panel. The dark and light shaded areas correspond to the regions enclosing 68\% and 95\% of the sets of solutions, respectively.}
    \label{f:DMU_ecc_distr}
\end{figure*}

Assuming a face-on orbit and treating Equation \ref{eq:dmu} as an equality, we can then find the pairs of mass $M_2$ and semi-major axis $a$ of secondary companions that could produce the observed $\Delta\mu$ of the primary for these $\mathcal{R}_0$ upper and lower limits. This provides the boundaries of the region in the $M_2-a$ space where the unseen companion inducing the observed change in proper motion could be, for a fixed eccentricity $e$ for the system.
In the right panel of Figure \ref{f:DMU_fixed_ecc}, we present the output of the code for an example target of mass $M_1 = 1$ M$_\odot$ with a parallax $\varpi = 50$ mas (distance of 20 pc) and a supposed change in proper motion of $\Delta\mu = 5\pm1$ mas yr$^{-1}$, for the same eccentricity values as in the left panel.
For each eccentricity, the $M_2-a$ pairs of mass and semi-major axis of companions compatible with the $\Delta\mu$ value correspond to the parameter space enclosed within the shaded areas for that $e$ value. The compatible range of $M_2-a$ pairs steadily broadens as the eccentricity increases and more $\mathcal{R}_0$ values are allowed, as a direct result of the wider ranges of possible $\mathcal{R}_0$ values in the left panel. Assuming a circular orbit results in a single line of possible solutions (black solid line), since $\mathcal{R}_0$ is constant for $e=0$. For inclined orbits, the derived lines of solutions provide lower mass limits at each separation. We note that the proper motion offset $\Delta\mu$ is considered to correspond to the astrometry of primary star only, rather than the photocentre of an unresolved system. This effect is further discussed in Section \ref{catalogue_length} and will be negligible in cases of low-mass, faint secondaries such as those sought here.

Adopting a plausible distribution of eccentricities of the user's choice, we can now apply the same approach to obtain a single distribution of mass-separation pairs for the companion causing the proper motion offset observed on the primary. To take into account the uncertainty in the astrometric data, we generate $10^5$ $\Delta\mu$ values drawn from a Gaussian distribution centred on the measured value, with a standard deviation set to the error on the measurement. For each $\Delta\mu$ value, we then draw an eccentricity $e$ from the adopted distribution. In order to marginalise over all possible orbital phases, we randomly draw a value between 0 and 2$\pi$ for the mean anomaly $\mathcal{M}$ (equivalent to picking a random fraction of the orbital period from a uniform distribution), and compute the corresponding $\mathcal{R}_0$ for the considered $e$ value. Finally, we can find the pairs of masses and semi-major axes that produce the drawn $\Delta\mu$ for the resulting orbital phase factor $\mathcal{R}_0$.

This provides us with a collection of $10^5$ sets of solutions, from which we can obtain a 2-dimensional map of the resulting distribution in the mass-separation space. The obtained results are based entirely on dynamical arguments, although we note that the secondary masses are technically computed relative to the adopted primary mass, which will generally be model-dependent. Figure \ref{f:DMU_ecc_distr} shows the output of the code for the model target considered above, adopting a uniform distribution in $e$ between 0 and 1 in the left panel \citep{Raghavan2010}, and a Gaussian distribution of mean $e=0$ and width 0.3, truncated to the range $0 \leq e < 1$, in the right panel \citep{Bonavita2013}. The thick lines indicate the positions of the mean (solid line) and median (dashed line) mass at each separation value. The shaded areas mark the limits of the 68\% (dark purple) and 95\% (light purple) confidence intervals. We adopt a highest probability density approach to compute the confidence levels, which provides the shortest interval enclosing a fraction $\alpha$ of the output distribution (see \citealp{Fontanive2018}). Again, the results presented in Figure \ref{f:DMU_ecc_distr} were computed assuming the that observed systems are in face-on orbital configurations. The uncertainties introduced by such an assumption are investigated and discussed in Section \ref{inclination}.

As expected from Figure \ref{f:DMU_fixed_ecc}, assuming a uniform distribution in eccentricity results in a much broader output (left panel of Figure \ref{f:DMU_ecc_distr}) than for a distribution concentrated around low eccentricity values (right panel). We note that in both cases the median (dashed black line) is very close to the $e=0$ line of solutions in Figure \ref{f:DMU_fixed_ecc}. As a wider range of eccentricities is allowed, the mean of the resulting solutions shifts to higher masses at the same separations, as a result of the highly asymmetric intervals around the $e=0$ solution for high eccentricities (see Figure \ref{f:DMU_fixed_ecc}; note that the plots are all shown on logarithmic scales). The uniform eccentricity distribution is likely to be more accurate for companions in the stellar mass range, consistent with the roughly flat $e$ distribution observed by \citet{Raghavan2010} for stellar binaries with mass ratios larger than $q\sim0.1$.
On the other hand, the truncated Gaussian distribution considered here is thought to be more representative of the substellar and planetary population (see \citealp{Bonavita2013} and references therein). This normal distribution, restricted to the range $e$ = 0$-$1, is also very close to the Beta distribution derived by \citet{Kipping2013} for the exoplanet eccentricity distribution, and is thus in good agreement with the observed properties of known exoplanets, at least on short orbital separations.

\subsection{Promising candidates for direct imaging searches}
\label{promising_candidates}

Based on stellar proper motions from various astrometric catalogues, we can select stars showing significant discrepancies (e.g. $>$ 3$\sigma$) between short-term proper motions (i.e. \textit{Gaia} DR2, \textit{Hipparcos}) and measurements acquired over longer time baselines (e.g. TGAS, Tycho-2).
Using the output from the previous section to evaluate the characteristics of possible hidden companions, our new COPAINS tool can be used to assess whether a selected $\Delta\mu$ star might be a promising target in a direct imaging search for low-mass companions. 
The obtained results may not be fully accurate depending on the periods of the systems and the proper motion measurements considered (see Section \ref{limitations}). Nevertheless, the output of the code provides a fairly good idea of the region of the parameter space where the hidden companion might be located, given the assumptions made, and can be used to identify the most interesting targets for a direct imaging campaign. Comparing the resulting 2-dimensional distribution of masses and separations to typical performances of imaging facilities, we can select stars for which the achieved sensitivities specifically probe the part of the parameter-space corresponding to substellar or planetary companions in our dynamical predictions. 
We can also use detection limits from direct imaging data with null detections to rule out a range of solutions and further constrain the region of the parameter space in which the unseen companion may be located.

\begin{figure}
    \centering
    \includegraphics[width=0.47\textwidth]{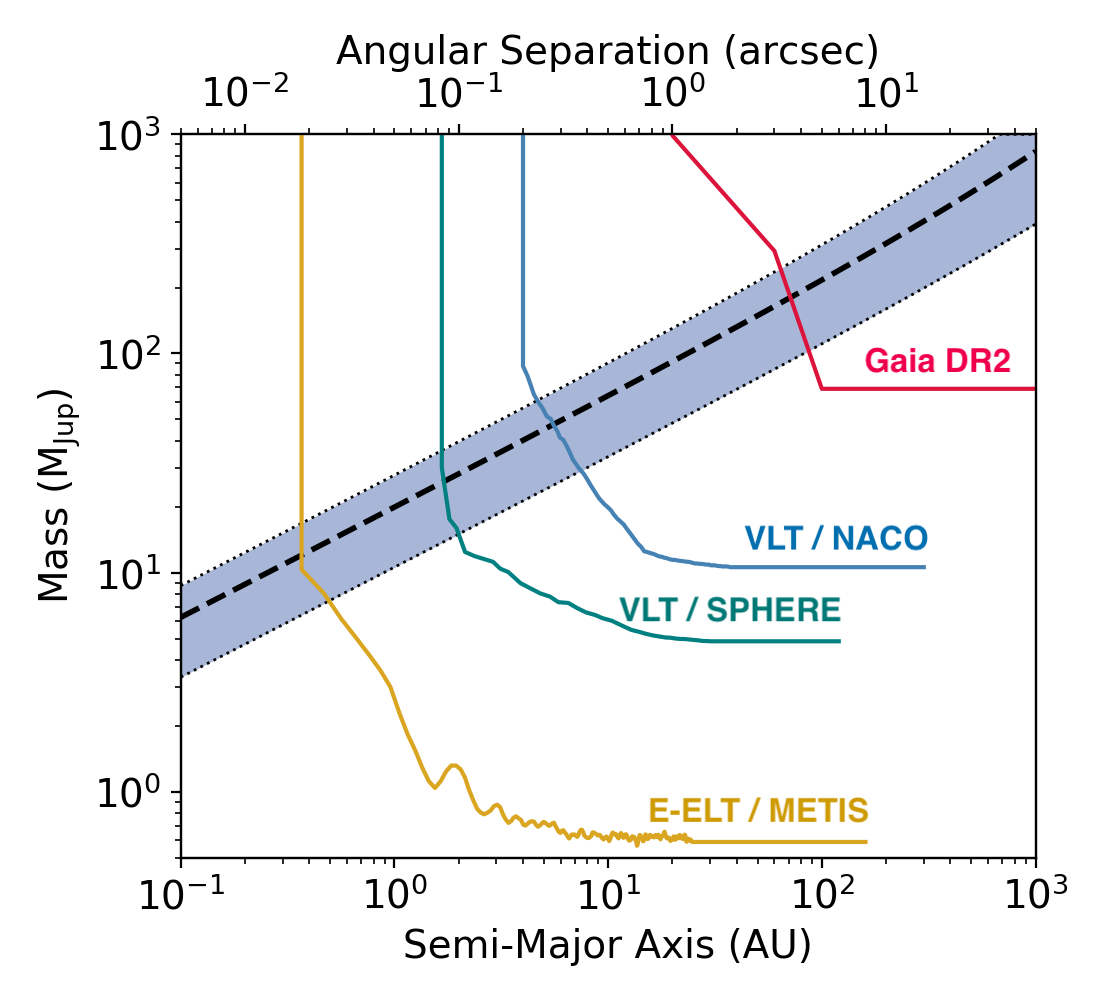}
    \caption{Output of COPAINS for our example target, assuming a flat distribution in eccentricity (see left panel of Figure \ref{f:DMU_ecc_distr}), showing the median (dashed black like) and 1-$\sigma$ interval (shaded region) of the possible solutions for the mass and separation of the astrometric companion. We plot the typical sensitivity limits to secondaries from \textit{Gaia} DR2 (red), VLT/NACO (blue), VLT/SPHERE (green) and E-ELT/METIS (yellow) as described in the text. Detection limits were converted to masses using the AMES-Cond evolutionary models \citep{Baraffe2003} and adopting an age of 100 Myr for our model target.} 
    \label{f:DMU_with_limits}
\end{figure}

For example, Figure \ref{f:DMU_with_limits} shows the results obtained with COPAINS on our example target from Section \ref{computation}, in which we overplotted typical sensitivities from various instruments. The VLT/NACO ($L'$ filter) and VLT/SPHERE ($YJH+K1K2$) 5-$\sigma$ detection limits were taken from the median performances achieved in \citet{Rameau2013} and \citet{Maire2017}, respectively. The \textit{Gaia} limit corresponds to the \textit{Gaia} DR2 99\% binary completeness limit described in \citet{Fontanive2019}. The sensitivity limits in terms of contrast for the future E-ELT/METIS instrument were obtained through end-to-end simulations of a 1-hour observation in the $L$ band for a star of magnitude $L=6$, using the classical vortex coronagraph and assuming an ADI sequence with 40 deg parallactic angle rotation (B. Carlomagno et al., in prep.). These sensitivity limits take into account residual wavefront errors from the adaptive optics system \citep{Hippler2019}, as well as shot noise from the star and thermal background.

Magnitudes were converted into masses using the AMES-Cond evolutionary models \citep{Baraffe2003} in the corresponding filter systems adopting an age of 100 Myr for our 1 M$_\odot$ target located at 20 pc. These values are representative of stars observed in direct imaging surveys, which typically target young and nearby objects to increase the detectability of low-mass, faint companions \citep{Vigan2017}. As the models do not include the METIS filter system yet, we used the NACO $L'$ bandbass to convert the METIS limits, since the METIS broadband $L$ filter is currently designed to be a copy of the NACO $L'$ filter.

The absence of comoving sources around this star in the \textit{Gaia} DR2 catalogue would allow us to rule out the part of the parameter space above the \textit{Gaia} limit (red line), eliminating low-mass stellar companions from $\sim$100 AU. The positions of the detection limits from the two VLT instruments then suggest that an intermediate-mass brown dwarf would be detectable with NACO (blue line), down to separations of around 10 AU. On the other hand, observations with SPHERE (green line) would be required to retrieve a hidden companion at the low-mass end of the substellar regime, on separations of a few AU. We note that the angular separation axis corresponds to the semi-major axis of the system, while the detection limits are technically in terms of projected separations, which would be expected to be slightly offset due to orbital eccentricity effects.

Finally, we can also use the plot in Figure \ref{f:DMU_with_limits} to conclude that a combination of VLT/SPHERE data with a binary search in \textit{Gaia} DR2 should allow for the retrieval of the unseen companion given that its separation is larger than $\sim$2 AU. A null detection in both data sets would place robust constraints on the mass and separation upper limits of the companion triggering the observed astrometric trend of the host star. A complementary detection method such as radial velocity (if the orbital inclination of the system is not face-on), or observations with next-generation facilities like the E-ELT/METIS instrument, could then be used to detect a very low-mass companion that remains undetected with current imaging capabilities.

\section{Assessment of the limitations of the method}
\label{limitations}

\subsection{Effect of duration of catalogues}
\label{catalogue_length}

The accuracy of the results obtained with this method strongly depends on how close the long-term proper motion measurement is to the true motion of the system's barycentre, and how close the short-term proper motion is to the instantaneous velocity due to the reflex orbital motion of the primary. If the period of the binary is too short, the short-term proper motion will cover a relatively large fraction of the orbit and be highly inconsistent with the tangential instantaneous velocity at the mean epoch of the proper motion observations. Similarly, in binary systems with very long periods, long-term proper motions may still capture some of the orbital motion of the primary and differ significantly from the true barycentric motion.

In order to test the effect of the offsets introduced by using catalogues of various baselines, we estimated the difference between the proper motion that would be observed over different time baselines, and the true instantaneous or centre-of-mass displacements, for a range of companions around an example target of mass $M_1 = 1$ M$_\odot$. We used a parallax of $\varpi = 50$ mas and assumed a proper motion of $\mu_{\alpha*} = 100$ mas yr$^{-1}$ and $\mu_\delta = 100$ mas yr$^{-1}$ for the barycentre of the system. We considered the \textit{Gaia} DR2 and \textit{Hipparcos} catalogues for short-term proper motions, and the TGAS and Tycho-2 catalogues for long-term measurements.

We first constructed a grid of semi-major axes $a$ and companion masses $M_2$, with 100 log-spaced $a$ values between 0.1$-$1000 AU, and 100 log-spaced $M_2$ in the range 0.001$-$1 M$_\odot$. For each cell in the grid, we generated 1000 random orbits by drawing from uniform distributions an eccentricity $e$, argument of periastron $\omega$, longitude of ascending node $\Omega$ and time of periastron passage $T_0$, and adopting a uniform distribution in $\sin{i}$ for the inclination $i$, in order to generate random orientations of orbits in space. Given the adopted system parameters and the drawn orbital elements, we can compute the on-sky trajectory of the primary for each simulated orbit following the standard formalism involving the Thiele-Innes elements (\citealp{Thiele1883}; see e.g. \citealp{Heintz1978} for details).

We then derived measured proper motions by considering the position of the primary at the start and end observation dates of each catalogue, and estimating the velocity vector between these two points given the length of the considered catalogue. For \textit{Gaia} DR2, proper motions were calculated between 2014.6 and 2016.4, and were compared to instantaneous velocities at epoch 2015.5 \citep{GaiaCollaboration2018}. For \textit{Hipparcos}, we considered observing dates between 1989.8 to 1993.2 \citep{Perryman1997}, with a mean epoch of 1991.5 for the tangential velocity, a bit later than the catalogue epoch (1991.25; \citealp{ESA1997}). Proper motions were estimated between 1900 and 2000 for Tycho-2, as the original epochs used to derive Tycho-2 proper motions from 144 ground-based photographic programs vary by a few decades around this initial epoch \citep{Hog2000}.
Finally, we used observation dates between 1991.25 and 2015.0 for TGAS, the epochs of the positional measurements in Tycho-2 and \textit{Gaia} DR1 \citep{Michalik2015}.

For short-term proper motions, we computed the instantaneous velocity of the primary at the mean epoch of the short-term catalogue. For long-term catalogues, the obtained proper motion measurement was compared to the adopted true motion of the system given above. For each simulated orbit, we then computed the modulus of the vector difference between a short or long term-proper motion measurement and the true underlying tangential velocity or centre-of-mass motion. Finally, we calculated the upper limit of the 1-$\sigma$ interval of the 1000 resulting differences to obtain a single value for each cell in the original grid. The 1-$\sigma$ confidence level was found by adopting a highest probability density approach, which provides the set of most probable values (see Section \ref{computation}). We considered the upper boundary of the 68\% credible interval rather than the mean or median of the samples in each cell of the grid in order to take into account the spread of the obtained distributions, which may vary throughout the parameter space. The final values thus provide uniformly-defined upper limits on the expected disparities, and only a small fraction of systems in the remaining low-probability upper tails are expected to be above these thresholds.

\begin{figure*}
    \centering
    \begin{minipage}[b]{0.495\textwidth}
        \includegraphics[width=\textwidth]{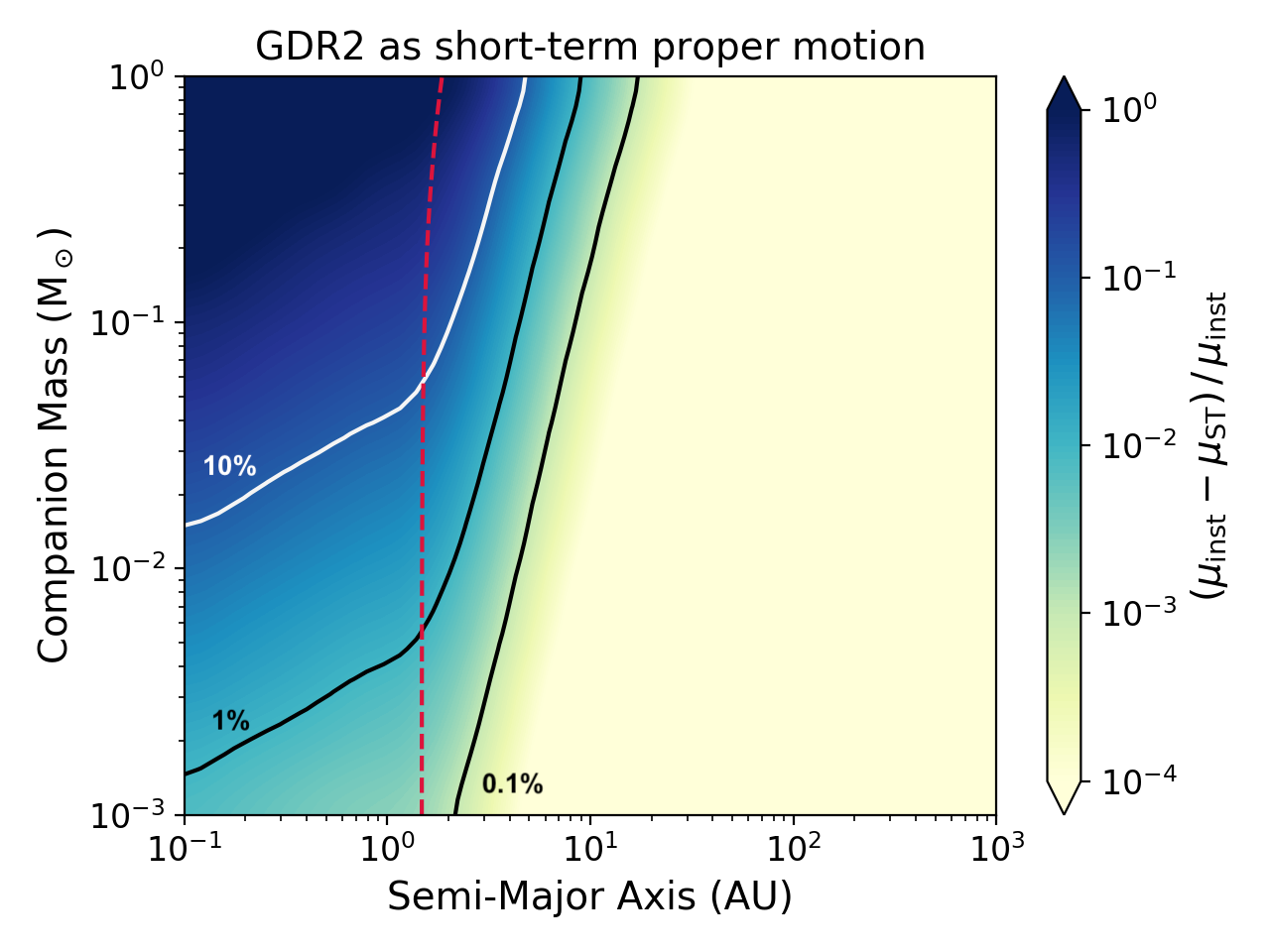}
    \end{minipage}
    \begin{minipage}[b]{0.495\textwidth}
        \includegraphics[width=\textwidth]{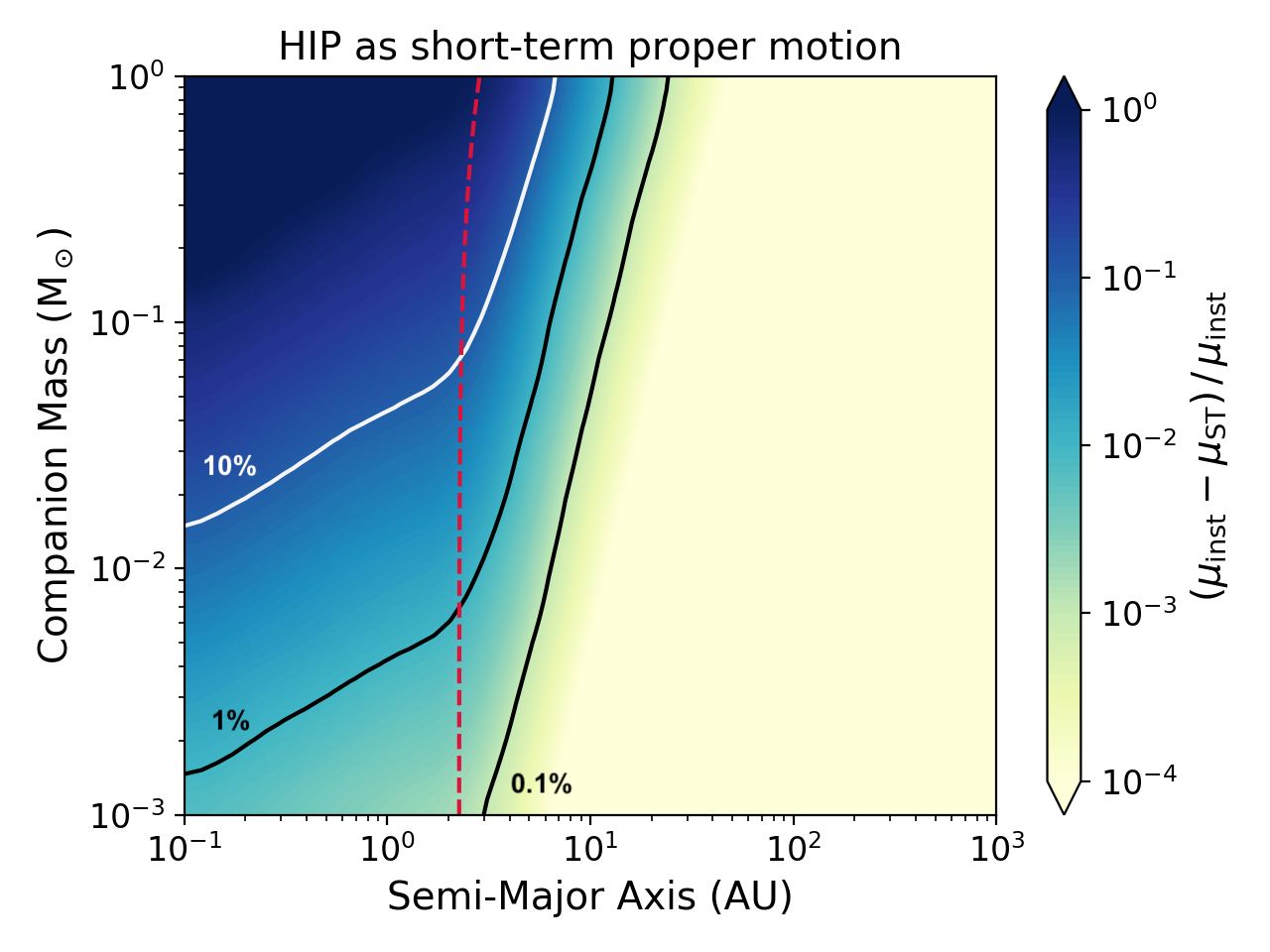}
    \end{minipage}
    \begin{minipage}[b]{0.495\textwidth}
        \includegraphics[width=\textwidth]{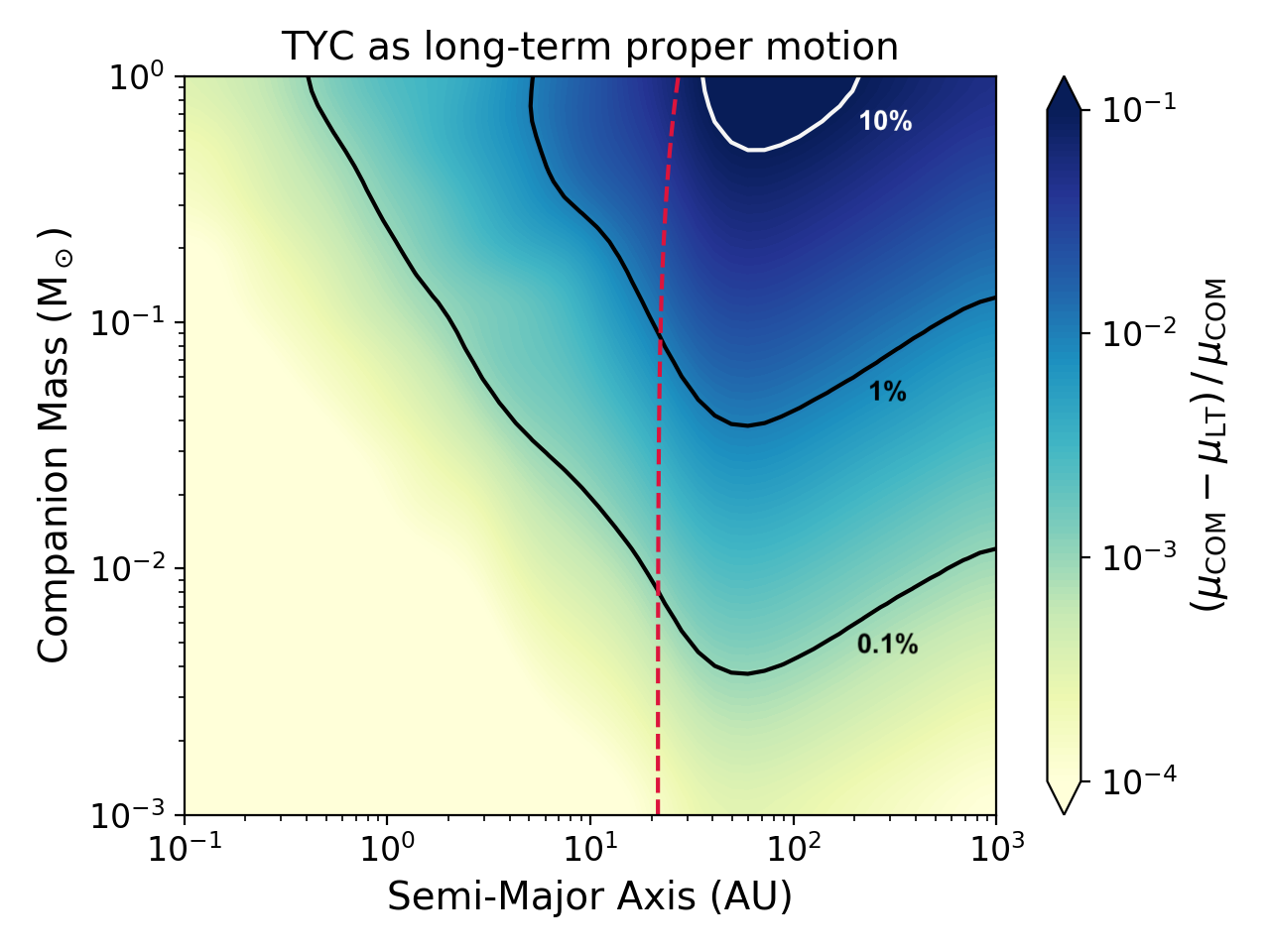}
    \end{minipage} 
    \begin{minipage}[b]{0.495\textwidth}
        \includegraphics[width=\textwidth]{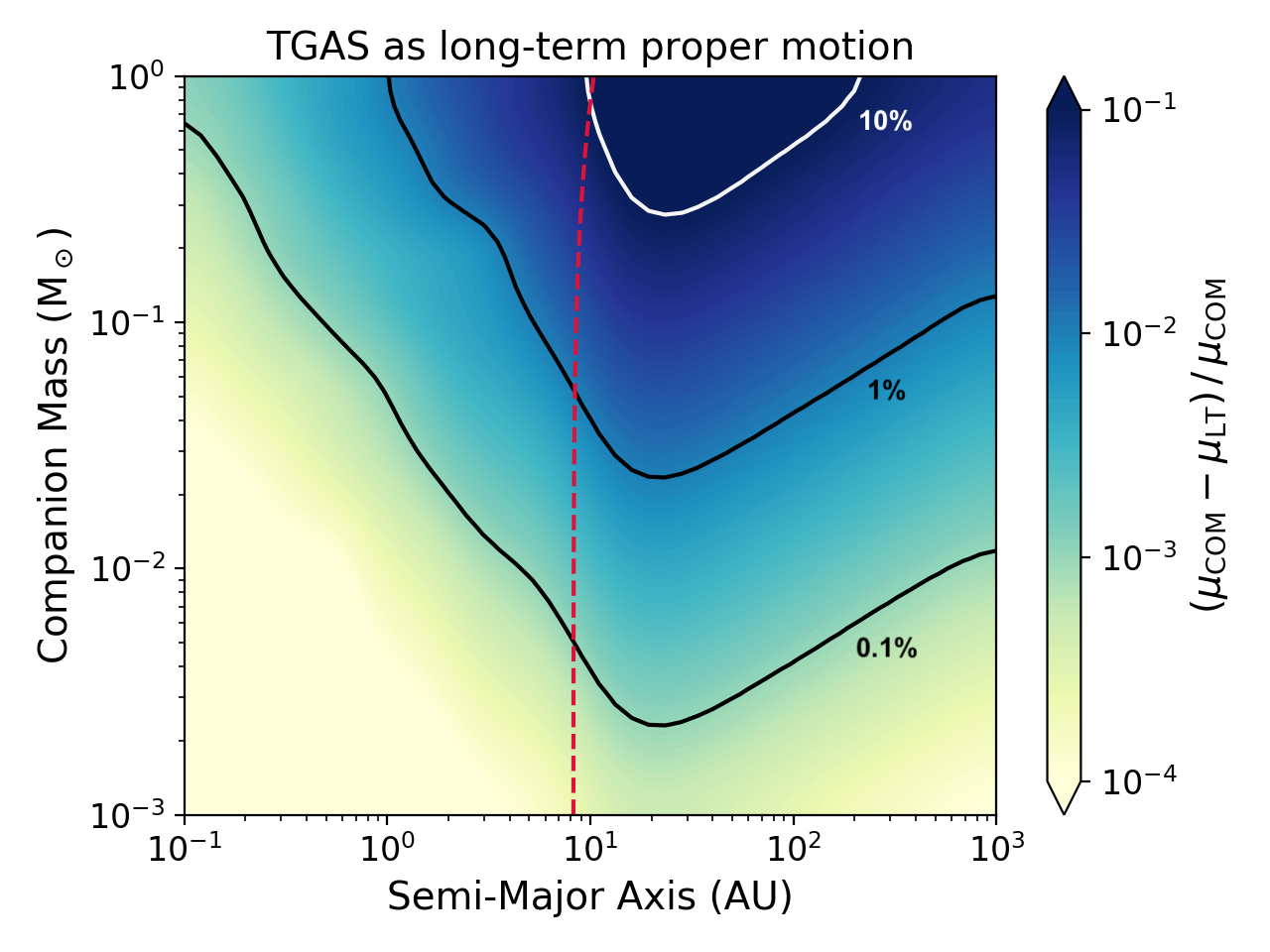}
    \end{minipage}
    \caption[Relative offset between measured short and long-term proper motions and instantaneous velocities and centre-of-mass motions.]{\textbf{Top:} 1-$\sigma$ upper boundary of the relative offset between the true instantaneous velocity of the primary and \textit{Gaia} DR2 (left) or \textit{Hipparcos} (right) proper motion measurements. The results are marginalised over all possible orbital elements for each pair of semi-major axis and companion mass around our 1 M$_\odot$ example target at 20 pc ($\varpi = 50$ mas). We adopted a proper motion of $\mu_{\alpha*} = 100$ mas yr$^{-1}$ and $\mu_{\alpha*} = 100$ mas yr$^{-1}$ for the model target. Values correspond to the upper limit of the 68\% interval of highest probability computed for 1000 randomly-selected orbital configurations for each mass-separation pair in the grid (see text). The red dotted lines indicate the mass-separation pairs for which the orbital periods of the systems are equal to the length of the catalogue (1.8 yr and 3.4 yr for \textit{Gaia} DR2 and \textit{Hipparcos}, respectively).
    \textbf{Bottom:} 1-$\sigma$ upper boundary of the relative difference between the barycentric motion of the system and long-term proper motion measurements of the primary from Tycho-2 (left) and TGAS (right). The red dotted lines show the position on the parameter space where systems have orbital periods equal to the duration of the long-term proper motion baselines (100 yr and 23.75 yr for \textit{Gaia} DR2 and \textit{Hipparcos}, respectively).}
    \label{f:offset}
\end{figure*}

The top panels of Figure \ref{f:offset} shows the resulting 1-$\sigma$ upper limits on the offsets on short-term proper motion measurements from the tangential velocities in the $a - M_2$ parameter space for \textit{Gaia} DR2 (left) and \textit{Hipparcos} (right) data. The solid lines show the regions where measured short-term proper motions are discrepant by 0.1\%, 1\% and 10\% from the instantaneous velocities we assume they represent.
As expected, high discrepancies arise in binaries of small separations, for which the short-term catalogues cover a significant fraction of the orbit (and up to multiple orbital periods). Large uncertainties will thus be introduced when using \textit{Gaia} DR2 or \textit{Hipparcos} as short-term proper motion measurements for systems with periods shorter than or comparable to the duration of the catalogues (red dotted lines). Longer orbital periods are required for the offsets to become insignificant, typically of at least several to tens of AU, for systems similar to our example target. 

In the bottom panels of Figure \ref{f:offset}, we present the disparities between long-term proper motion estimates and the true system proper motion using Tycho-2 (left) and TGAS (right). As for the top panels, the colour scale corresponds to the upper boundaries of the 68\% confidence intervals of the obtained distribution in cell. In this case, short-period binaries are preferred for the approximations to be reliable, where the proper motion measurements encompass multiple orbital periods. The offsets get larger as the orbital periods approach the length of the catalogues (red dotted lines). The turn-over at wider semi-major axes can be explained by the fact that for the same companion mass, the wobble induced on the primary becomes smaller with larger separations (Equation \ref{eq:dmu}). As a result, the proper motion of the primary generally remains closer to the barycentric motion, and the typical offset relative to the same true system motion decreases, even if the baseline of the catalogue only covers part of the orbital period. For our example star, the discrepancies are negligible for most substellar companions within $\sim$10 AU for Tycho-2 and a few AU when using TGAS.

\begin{figure*}
    \centering
    \begin{minipage}[b]{0.495\textwidth}
        \includegraphics[width=\textwidth]{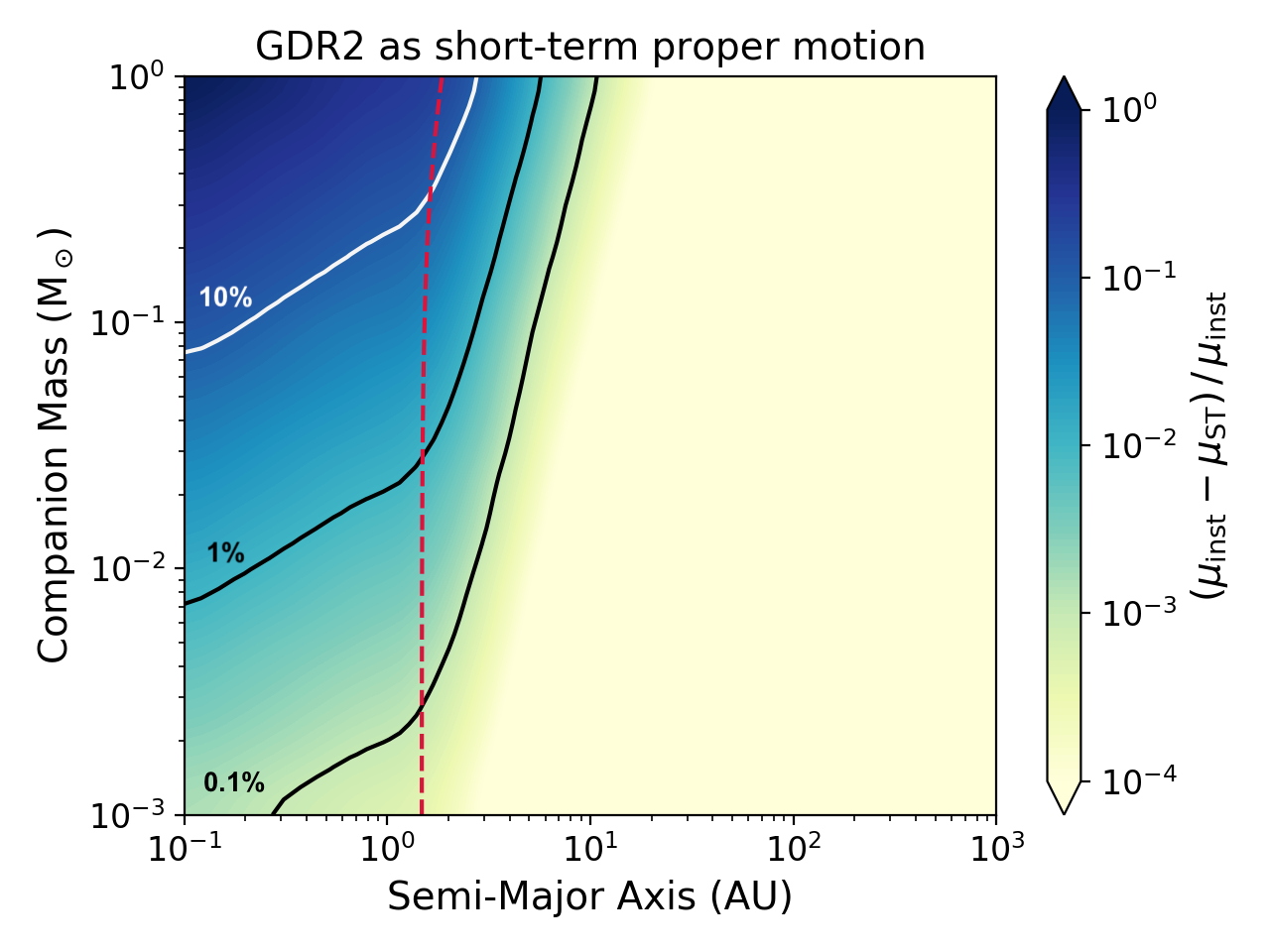}
    \end{minipage} 
    \begin{minipage}[b]{0.495\textwidth}
        \includegraphics[width=\textwidth]{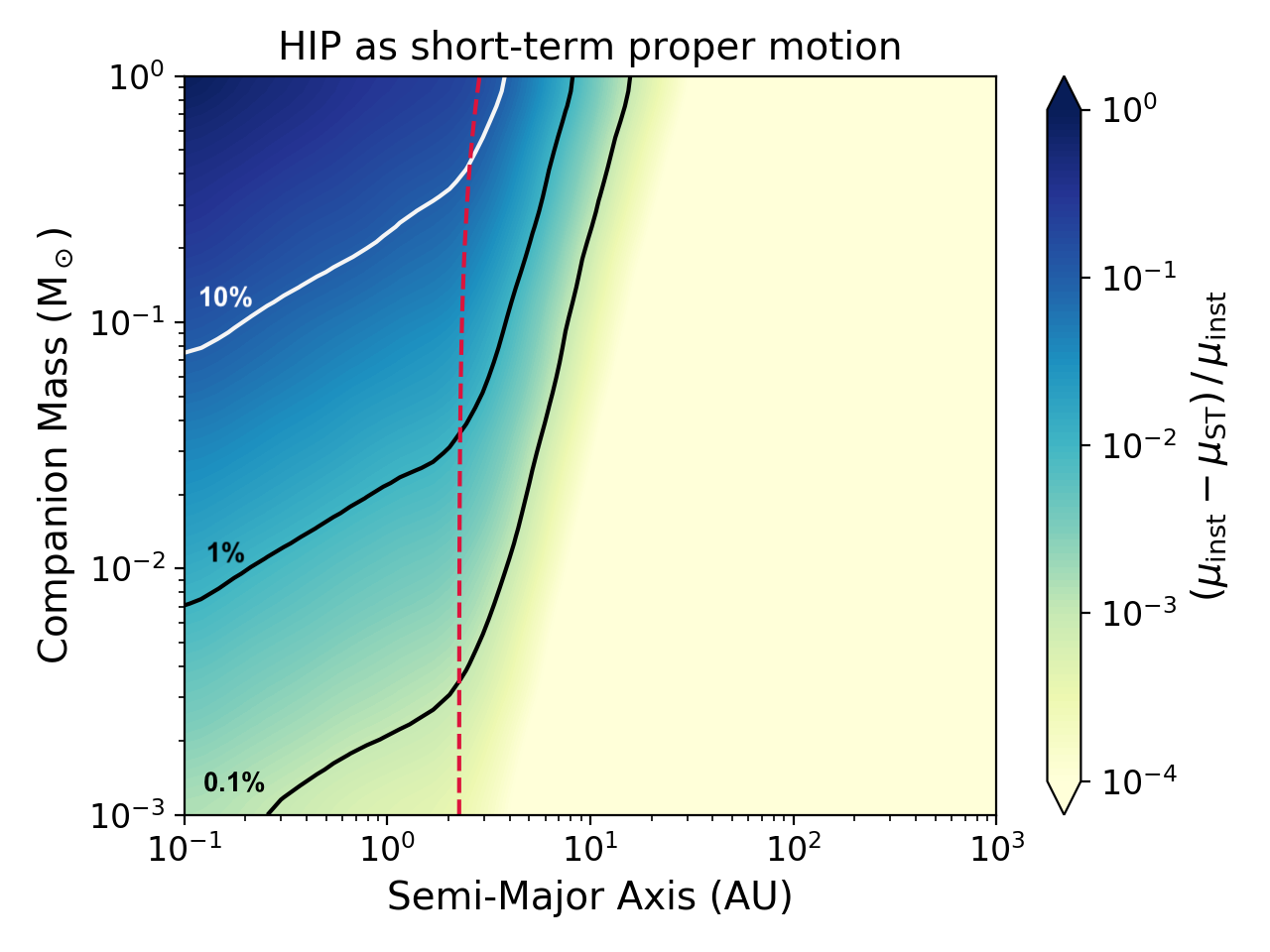}
    \end{minipage}
    \begin{minipage}[b]{0.495\textwidth}
        \includegraphics[width=\textwidth]{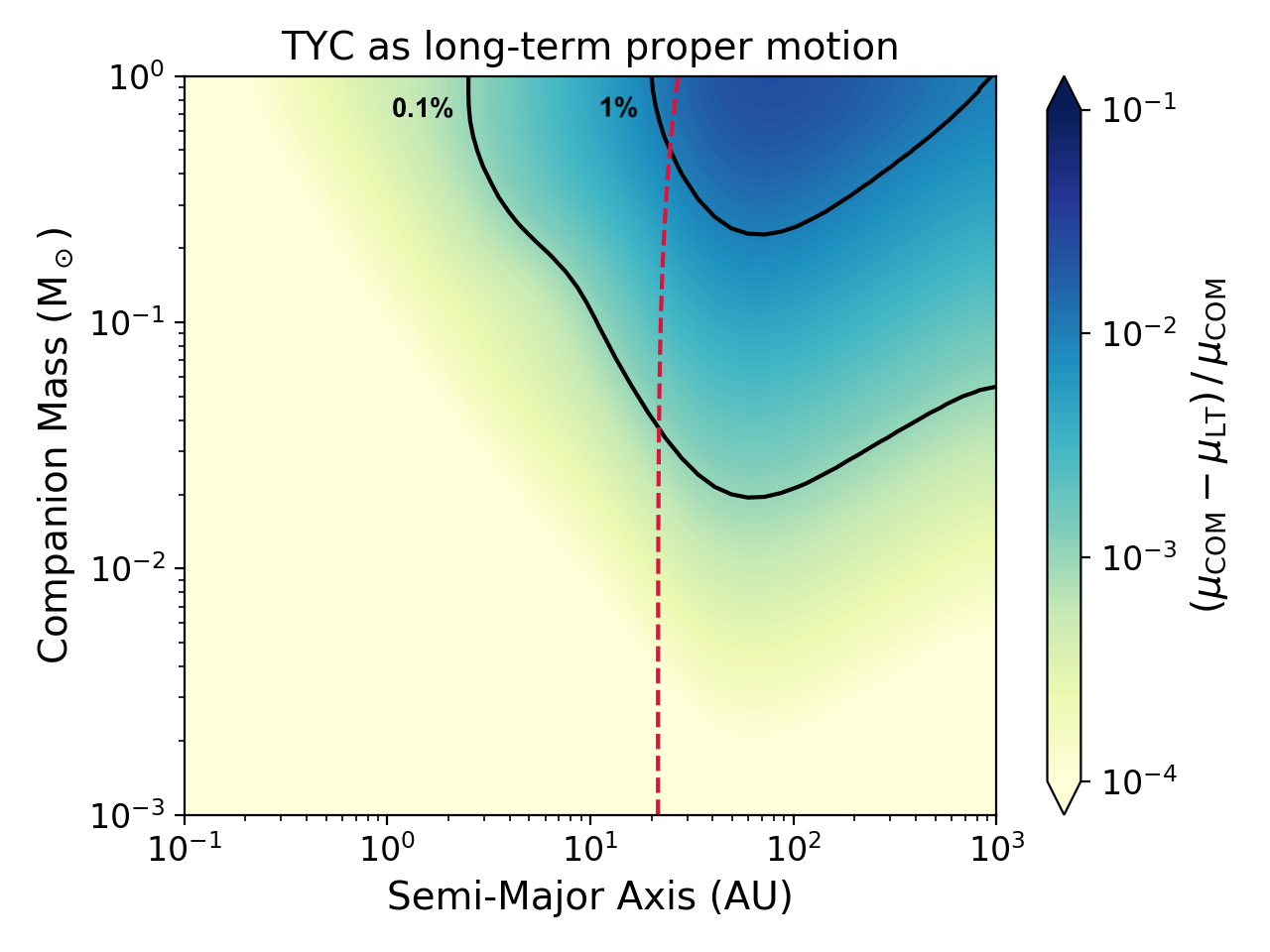}
    \end{minipage} 
    \begin{minipage}[b]{0.495\textwidth}
        \includegraphics[width=\textwidth]{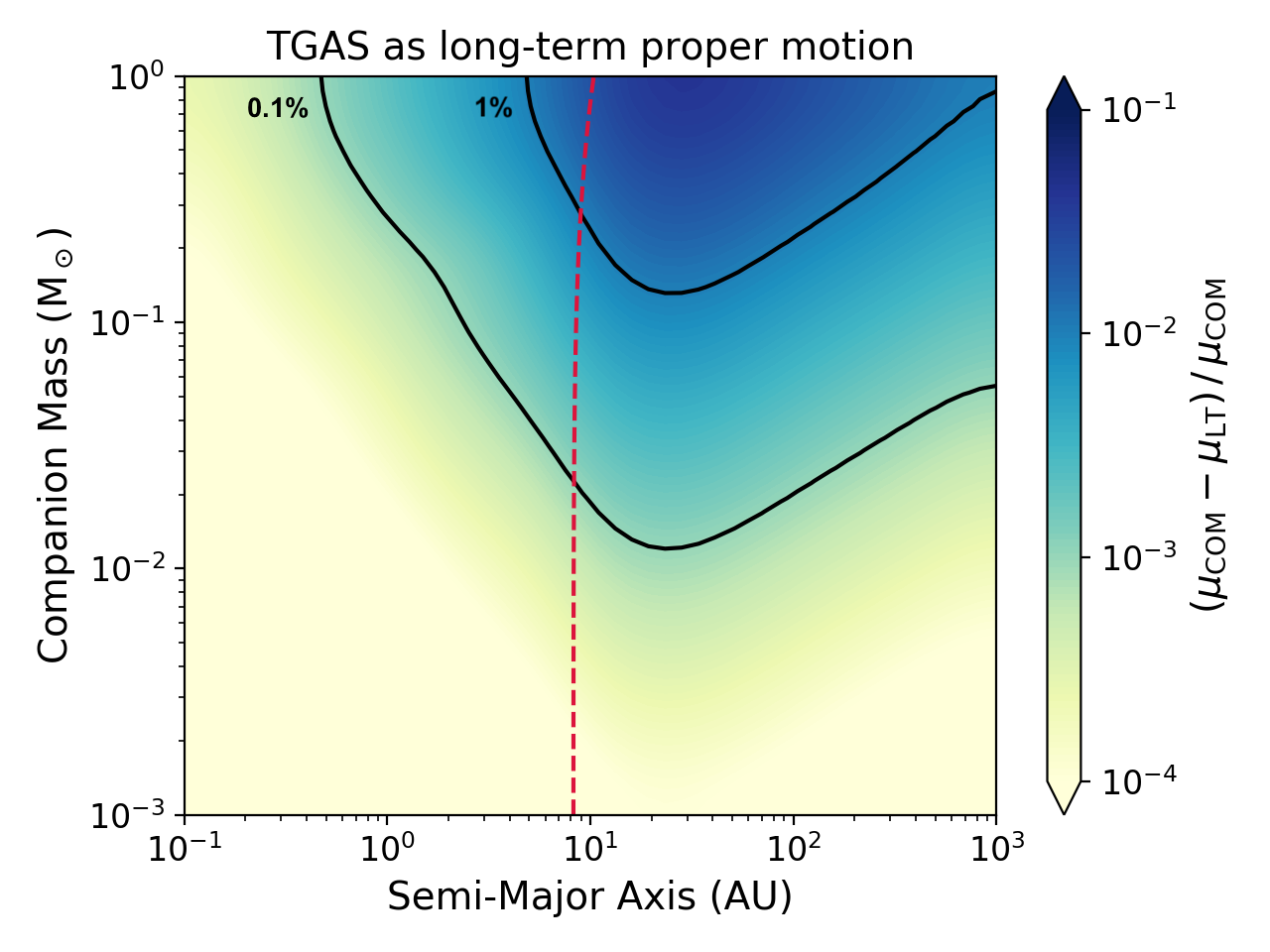}
    \end{minipage}
    \caption[Same as Figure \ref{f:offset} with decreased parallax.]{Same as Figure \ref{f:offset}, placing our example target of mass 1 M$_\odot$ at 100 pc ($\varpi = 50$ mas), with a proper motion of $\mu_{\alpha*} = 100$ mas yr$^{-1}$ and $\mu_{\delta} = 100$ mas yr$^{-1}$. The decrease in parallax leads to smaller fractional offsets for both the short-term proper motions (top panels) and long-term proper motions (bottom panels) relative to the instantaneous velocities and centre-of-mass motions we take them to represent, respectively.}
    \label{f:offset_plx}
\end{figure*}

\begin{figure*}
    \centering
    \begin{minipage}[b]{0.495\textwidth}
        \includegraphics[width=\textwidth]{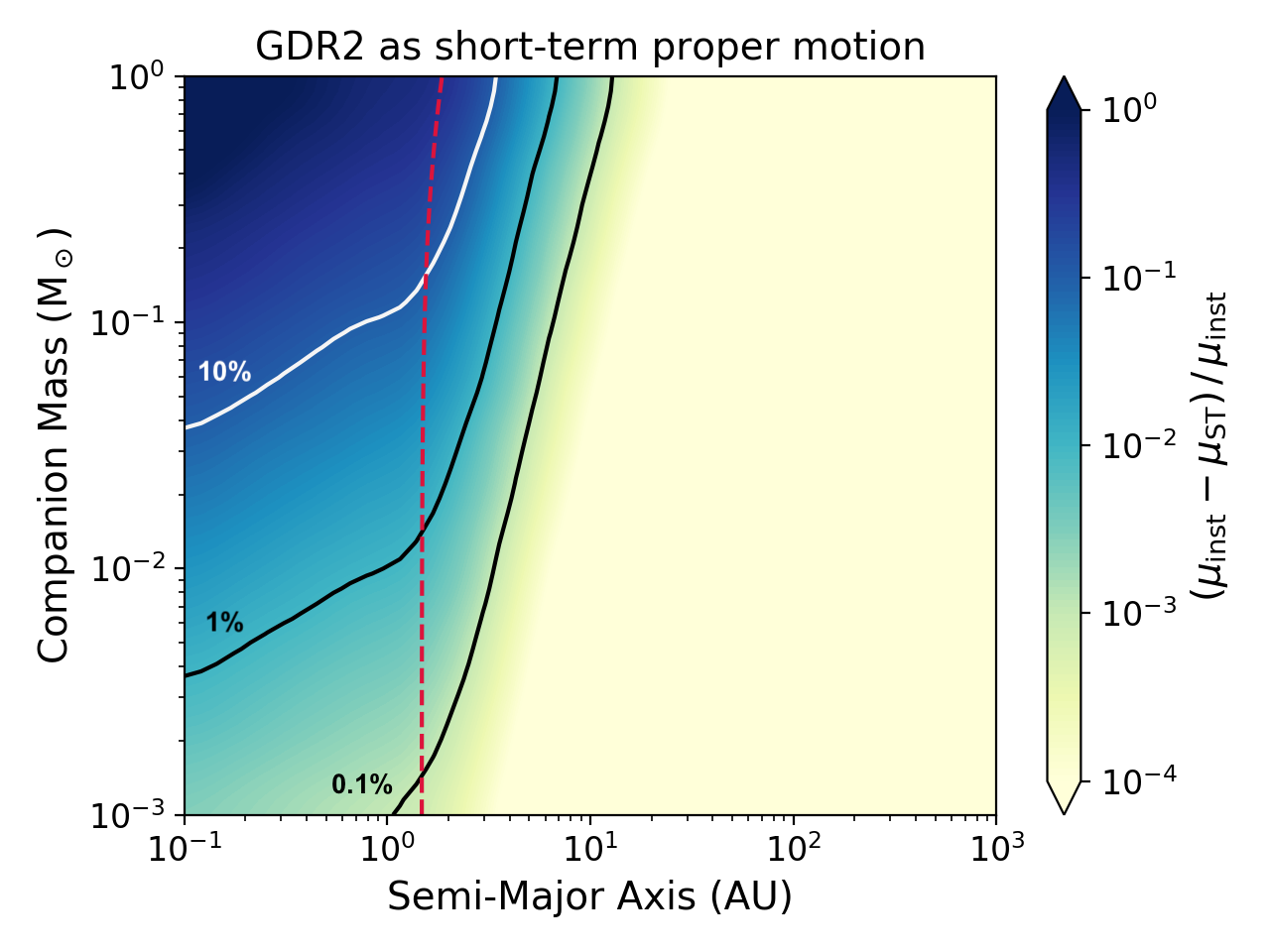}
    \end{minipage}
    \begin{minipage}[b]{0.495\textwidth}
        \includegraphics[width=\textwidth]{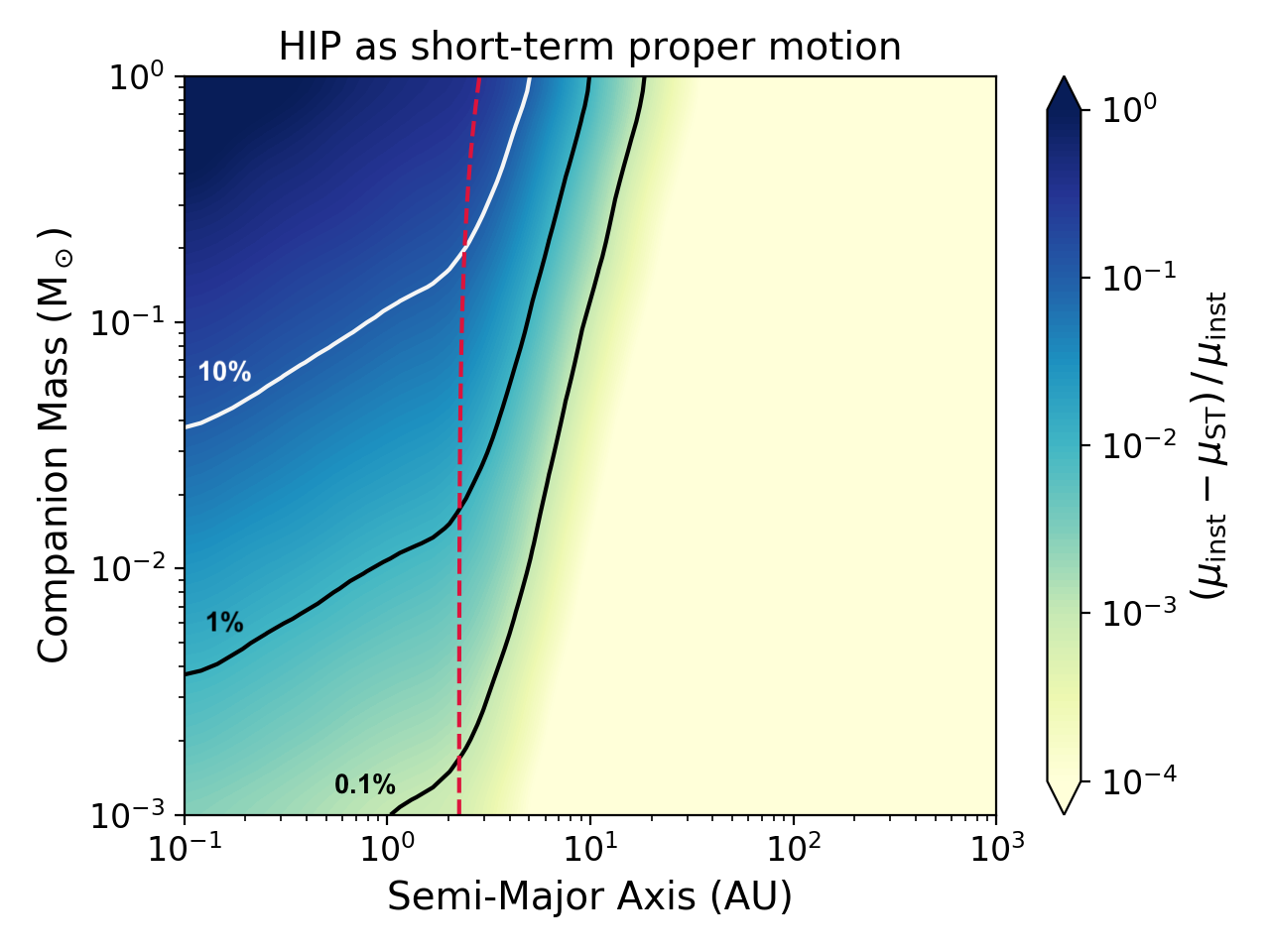}
    \end{minipage}
    \begin{minipage}[b]{0.495\textwidth}
        \includegraphics[width=\textwidth]{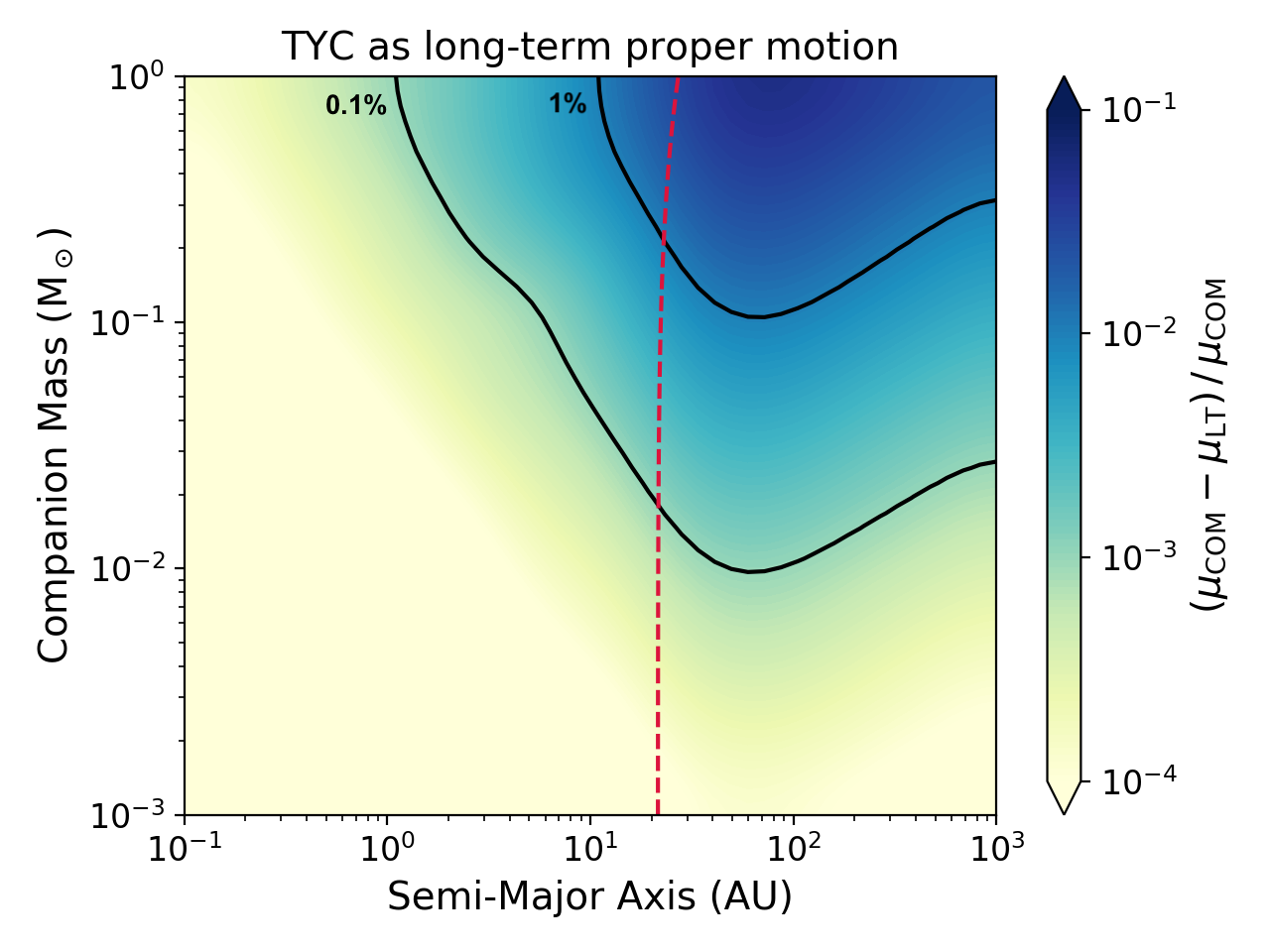}
    \end{minipage}
    \begin{minipage}[b]{0.495\textwidth}
        \includegraphics[width=\textwidth]{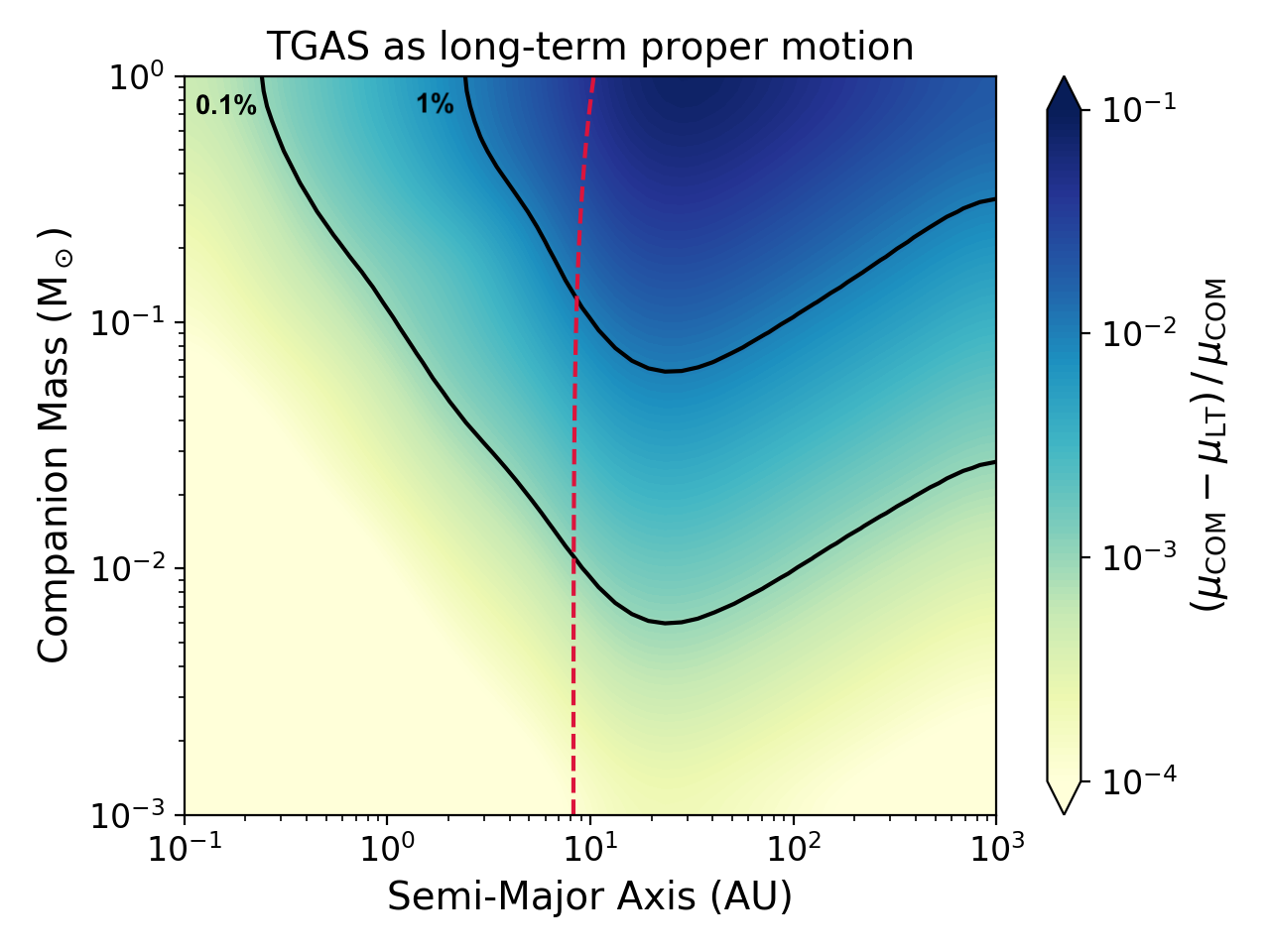}
    \end{minipage}
    \caption[Same as Figure \ref{f:offset} with increased proper motion.]{Same as Figure \ref{f:offset}, assigning a proper motion of $\mu_{\alpha*} = 250$ mas yr$^{-1}$ and $\mu_{\delta} = 250$ mas yr$^{-1}$ to our example target of mass 1 M$_\odot$ located at 20 pc ($\varpi = 50$ mas). The increase in proper motion magnitude results in smaller fractional offsets in both short-term (top panels) and long-term (bottom panels) proper motion measurements.}
    \label{f:offset_pm}
\end{figure*}

We note that as the obtained values represent upper limits at the 1-$\sigma$ level, our approach is rather conservative and these results are likely to be underestimating the usable regions of the parameter space, in which the relative offsets are small.
In addition, proper motions in the \textit{Hipparcos} and \textit{Gaia} DR2 catalogues are in reality acquired from many more positional measurements than the two data points at the beginning and end of the mission time spans \citep{vanLeeuwen2007,Lindegren2018}. Measured proper motions from these catalogues might therefore be somewhat closer to the true values than in our estimates. The uncertainties obtained here are thus likely to be overestimated for these reasons.

That being said, some of the proper motion information used might also carry further sources of uncertainties. For example, Tycho-2 astrometry may contain systematic errors from the archival data used to derive proper motions \citep{Michalik2015}. Astrometric 5-parameter solutions for \textit{Gaia} parallaxes and proper motions also assume that all stars are single \citep{Lindegren2018}. Since the $\Delta\mu$ targets of interest in this work are by definition not single, there may be some additional offset on the parallaxes and proper motions due to this effect, as a result of erroneously disentangling between parallax and proper motion on very short time spans \citep{Schonrich2019}.

We also investigated the dependence on distance in the results presented in Figure \ref{f:offset}. We found that at larger distances (smaller parallaxes), the relative disparities become smaller in every point of the parameter space, for both the short and long-term proper motion measurements. This is due to the fact that the observable excursion of the primary decreases with increasing distance for the same companion mass and semi-major axis. As a result, the obtained fractional offsets are smaller and the achieved relative accuracies are improved. The changes with decreasing parallax can be thought of as sliding the values in Figure \ref{f:offset} upwards at every semi-major axis, resulting in lower relative differences at any given mass between the measured proper motions and the underlying tangential or centre-of-mass velocities. Figure \ref{f:offset_plx} shows the same plots as in Figure \ref{f:offset}, placing our example target of 1 M$_\odot$ at a distance of 100 pc ($\varpi = 10$ mas) instead of 20 pc (50 mas), illustrating the upward shift in the obtained values relative to Figure \ref{f:offset}.

Similarly, we examined the effect of proper motion magnitude on the results obtained above. We set the proper motion of our example target to $\mu_{\alpha*} = 250$ mas yr$^{-1}$ and $\mu_{\delta} = 250$ mas yr$^{-1}$, for a parallax of $\varpi = 50$ mas (closer to the proper motions of the targets studied in Section \ref{application}). The same outcome was observed compared to decreasing the assigned parallax, as shown in Figure \ref{f:offset_pm}. Smaller relative offsets were obtained throughout the parameter space for larger system motions, in both the estimates of tangential velocities (top panels) and barycentric motions (bottom panel). Again, this is attributed to the fact that the same observable astrometric displacement of the primary due to the presence of a secondary companion will be subdued by the larger space motion of the system. As expected, the combination of larger parallax and higher proper motion results in a joint outcome of the two individual effects noted here. In this case, a further upwards shift of the obtained values in Figures \ref{f:offset} to \ref{f:offset_pm} is observed, decreasing the resulting uncertainties in every point in the parameter space.

Finally, the method also implicitly assumes that the astrometric data always correspond to observations of the primary component of a binary system. In reality, for unresolved binaries positional measurements reflect the space motion of the photocentre rather than the astrometry of the primary star itself. As the astrometric displacement of the photocentre is smaller than the excursion of the primary, the observed $\Delta\mu$ may thus be further underestimated \citep{Makarov2005}.

We conclude that in general, intermediate-separation systems with semi-major axes from $\sim$few AU up to several tens of AU have the lowest relative offsets in both short and long-term measurements compared to the instantaneous and barycentric motions, respectively.
The region of optimal coverage corresponds to the part of the parameter space in which most direct imaging companions are located. These results thus suggest that the $\Delta\mu$ from Equation \ref{eq:dmu} will be reasonably well approximated by, e.g., \textit{Gaia} DR2 and Tycho-2 proper motions for systems with orbital separations accessible with current imaging facilities.

\subsection{Effect of orbital inclination}
\label{inclination}

\begin{figure}
    \centering 
    \includegraphics[width=0.47\textwidth]{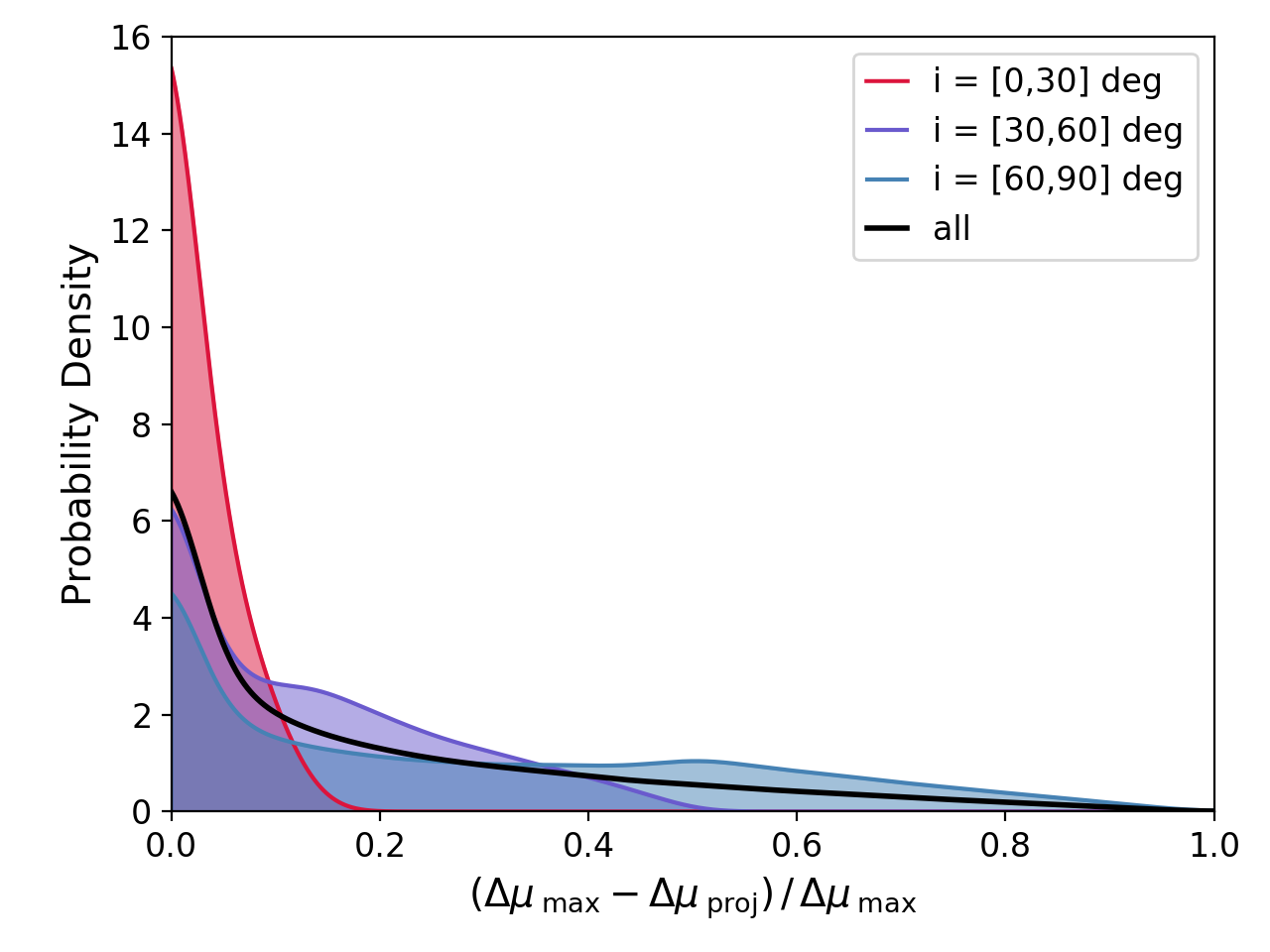}
    \caption[Distribution of relative offsets between projected $\Delta\mu$ measurements and the corresponding values for a face-on system.]{Distribution of relative offsets between projected $\Delta\mu$ measurements and the corresponding $\Delta\mu$ that would be observed if the same systems were in the plane of the sky (inclination $i = 0$ deg). The resulting disparities represent the error introduced in our approach by considering observed proper motion changes to correspond to face-on orbits (i.e. treating Equation \ref{eq:dmu} as an equality). The discrepancies were computed for 1000 randomly-selected orbits for each pair of semi-major axis and companion mass in a log-spaced grid of size $100\times100$ ranging from 0.1$-$1000 AU and 0.001$-$1 M$_\odot$. The solid black line represents the overall distribution, and the coloured segments correspond to various inclination bins.}
    \label{f:inclination_offset}
\end{figure}

Another source of uncertainty present in our procedure stems from the orbital inclination of binary systems. Equation \ref{eq:dmu} is treated as an equality to produce the plots in Figures \ref{f:DMU_ecc_distr} and \ref{f:DMU_with_limits}. The code therefore assumes face-on orbits and our method does not account for orbital inclination. As noted previously, in cases of inclined orbits a projected $\Delta\mu$ is observed, which can take values up to a maximum value corresponding to the face-on configuration. By assuming that the observed $\Delta\mu$ corresponds to this maximum value, the true maximum given by Equation \ref{eq:dmu} may be underestimated. The solutions obtained thus provide lower bounds on the region of the parameter space where the hidden companions might be (i.e. above and to the left of the resulting curves in Figures \ref{f:DMU_ecc_distr} and \ref{f:DMU_with_limits}).

We inspected the consequence of the assumption of face-on orbits in our approach. We considered the same example target as above ($M_1 = 1$ M$_\odot$, $\varpi = 50$ mas, $\mu_{\alpha*} = \mu_\delta = 100$ mas yr$^{-1}$). We used a similar logarithmic $a - M_2$ grid, with 100 semi-major axes from 0.1 to 1000 AU, and 100 companion masses ranging from 0.001 to 1 M$_\odot$. For every mass-separation pair, we generated 1000 random orbits (see Section \ref{catalogue_length}). For each simulated orbit, we then computed the projected proper motion offset $\Delta\mu_\mathrm{proj}$, given by the difference between the instantaneous velocity of the primary at epoch 2000 and the adopted centre-of-mass motion of the system stated above. This provides the proper motion difference that would be measured for the specific system considered.
We considered in parallel the same orbital elements with an inclination of $i = 0$ deg, and calculated the proper motion discrepancy of the same system and at the same epoch, assuming a face-on orbit. Finally, we estimated the relative difference between the obtained projected and maximum $\Delta\mu$ values: ($\Delta\mu_\mathrm{max} - \Delta\mu_\mathrm{proj}$) / $\Delta\mu_\mathrm{max}$. The obtained differences were normalised to $\Delta\mu_\mathrm{max}$ in order to be able to compare the output throughout the parameter space, as the induced changes in proper motion vary with companion mass and separation.

Figure \ref{f:inclination_offset} shows the distribution of relative disparities between the projected and face-on changes in proper motion for all 1000 simulated orbits in each cell in the mass-separation grid (black line). The results are also divided into three inclination bins, shown in the filled, coloured distributions. As expected, larger disparities were observed with higher inclination values. We note that the displayed inclinations bins are not weighted equally in the overall distribution (black line) due to the fact that we used a flat distribution in $\sin{i}$ rather than in $i$. In most configurations, the resulting offset is typically very small, with a clear peak near 0, and a median of 12\%. The relative differences were found to be smaller than 25\% at the 1-$\sigma$ level (68\% confidence interval) for the overall distribution. The normalised offsets between $\Delta\mu_\mathrm{proj}$ and $\Delta\mu_\mathrm{max}$ appeared to be distributed homogeneously throughout the parameter space.

We conclude that the assumption of a face-on orbit in our approach will typically lead to the $\Delta\mu$ values used in the code being underestimated by $<$ 15$-$20\%, and this offset will be negligible (less than $\sim$10\% disparity) in about half of all cases.
Given that most selected $\Delta\mu$ targets have a proper motion trend just above the 3-$\sigma$ selection threshold, this new source of uncertainty will generally be significantly smaller than the measurement uncertainty on the observed $\Delta\mu$, and the assumption of an orbit in the plane of the sky will not have a major effect on our results.

\section{Application of COPAINS to known directly-imaged systems}
\label{application}

We considered the five targets studied by \citet{Brandt2018} in order to validate the methodology of our tool (HD 4747, GJ 86, HD 68017, GJ 758 and HR 7672). The companions in all these systems have well-constrained orbits and dynamical mass estimates from combinations of radial velocity, direct imaging and astrometric data \citep{Brandt2018}, allowing us to robustly test the performance of COPAINS to select such systems.

\renewcommand{\arraystretch}{1.3}
\renewcommand{\tabcolsep}{0.4cm}
\begin{table*}
\centering
\caption{Properties and astrometry of the systems from \citet{Brandt2018}.}
\begin{tabular}{ l c c c c c }
\hline \hline
Property & HD 4747 & GJ 86 & HD 68017 & GJ 758 & HR 7672 \\
\hline
Primary mass $M_1$ (M$_\odot$) & $0.82^{+0.07}_{-0.08}$ & $1.36\pm0.23$ & $0.98\pm0.07$ & $0.76^{+0.13}_{-0.27}$ & $0.96^{+0.04}_{-0.05}$ \\
Secondary mass $M_2$ (M$_\mathrm{Jup}$) & $66.2^{+2.5}_{-3.0}$ & $623\pm11$ & $154\pm3$ & $38.1^{+1.7}_{-1.5}$ & $72.8\pm0.8$ \\
Semi-major axis $a$ (AU) & $10.1^{+0.4}_{-0.5}$ & $21.7^{+0.5}_{-0.7}$ & $16.0^{+1.0}_{-1.2}$ & $30^{+5}_{-8}$ & $19.6^{+0.8}_{-1.0}$ \\
Orbital period $P$ (yr) & $34^{+0.8}_{-1.0}$ & $72^{+7}_{-8}$ & $60^{+6}_{-8}$ & $180^{+60}_{-90}$ & $86^{+7}_{-8}$ \\
Eccentricity $e$ & $0.7353^{+0.0027}_{-0.0029}$ & $0.53^{+0.04}_{-0.03}$ & $0.325^{+0.017}_{-0.024}$ & $0.40\pm0.09$ & $0.542\pm0.018$ \\
Inclination $i$ (deg) & $49.4^{+2.3}_{-2.4}$ & $125.5^{+0.8}_{-0.9}$ & $170.3\pm0.4$ & $41\pm6$ & $97.4\pm0.4$ \\
Parallax $\varpi$ (mas) & $53.18\pm0.13$ & $92.70\pm0.05$ & $46.33\pm0.06$ & $64.06\pm0.02$ & $56.43\pm0.07$ \\
GDR2-TYC $\Delta\mu$ (mas/yr) & $0.95\pm1.45 ^*$ & $43.36\pm2.44$ & $19.93\pm0.81$ & $3.21\pm1.20$ & $12.91\pm1.02$ \\
HIP-TYC $\Delta\mu$ (mas/yr) & $5.57\pm1.60$ & $60.48\pm2.54$ & $5.01\pm1.37$ & $2.79\pm1.32$ & $1.96\pm1.24 ^*$ \\
%TGAS-TYC $\Delta\mu$ (mas/yr) & $3.38\pm1.4$ & $53.51\pm2.47$ & ... & $3.22\pm1.20$ & $4.87\pm1.00$ \\
GDR2-TGAS $\Delta\mu$ (mas/yr) & $3.29\pm0.35$ & $18.37\pm0.08$ & ... & $0.93\pm0.04$ & $8.45\pm0.09$ \\
HIP-TGAS $\Delta\mu$ (mas/yr) & $6.08\pm0.74$ & $18.74\pm0.52$ & ... & $1.75\pm0.52$ & $6.32\pm0.64$ \\
\hline \\[-0.3cm]
\multicolumn{6}{l}{
  \begin{minipage}{0.82\textwidth}
    \textbf{Notes.} Orbital elements come from \citet{Brandt2018}. Astrometric offsets were computed in this work based on proper motions taken directly from the \textit{Gaia} DR2 \citep{Lindegren2018}, TGAS \citep{Michalik2015}, \textit{Hipparcos} \citep{vanLeeuwen2007} and Tycho-2 \citep{Hog2000} catalogues, when available. Proper motion changes marked with a * have a significance $<$ 2$\sigma$. Parallax measurements correspond to \textit{Gaia} DR2 parallaxes.\\
  \end{minipage}}
\end{tabular}
\label{t:targets_table}
\end{table*}

\afterpage{
\begin{figure*}
    \centering
    \begin{minipage}[t]{0.445\textwidth}
        \includegraphics[width=\textwidth]{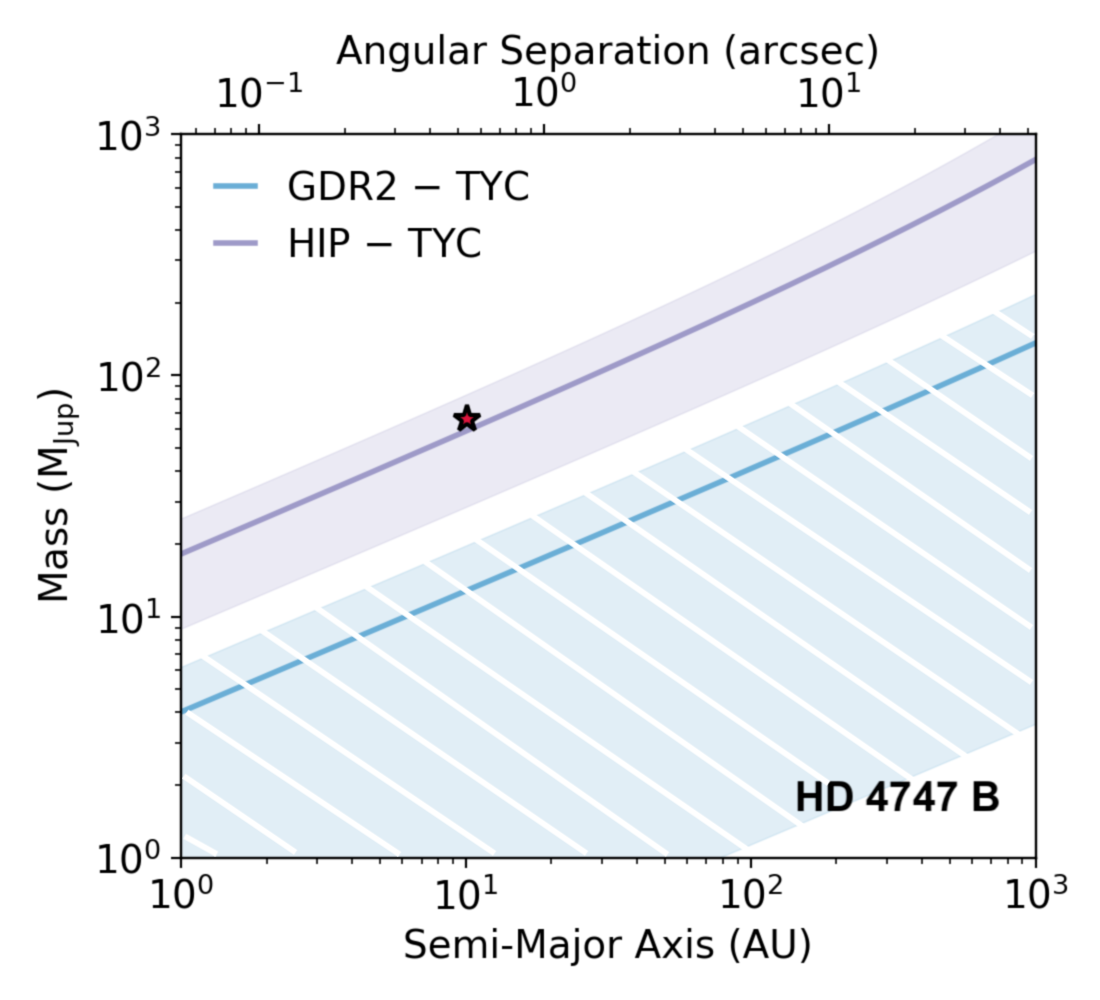}
    \end{minipage} \hspace{0.8cm} %\hfill
    \begin{minipage}[t]{0.445\textwidth}
        \includegraphics[width=\textwidth]{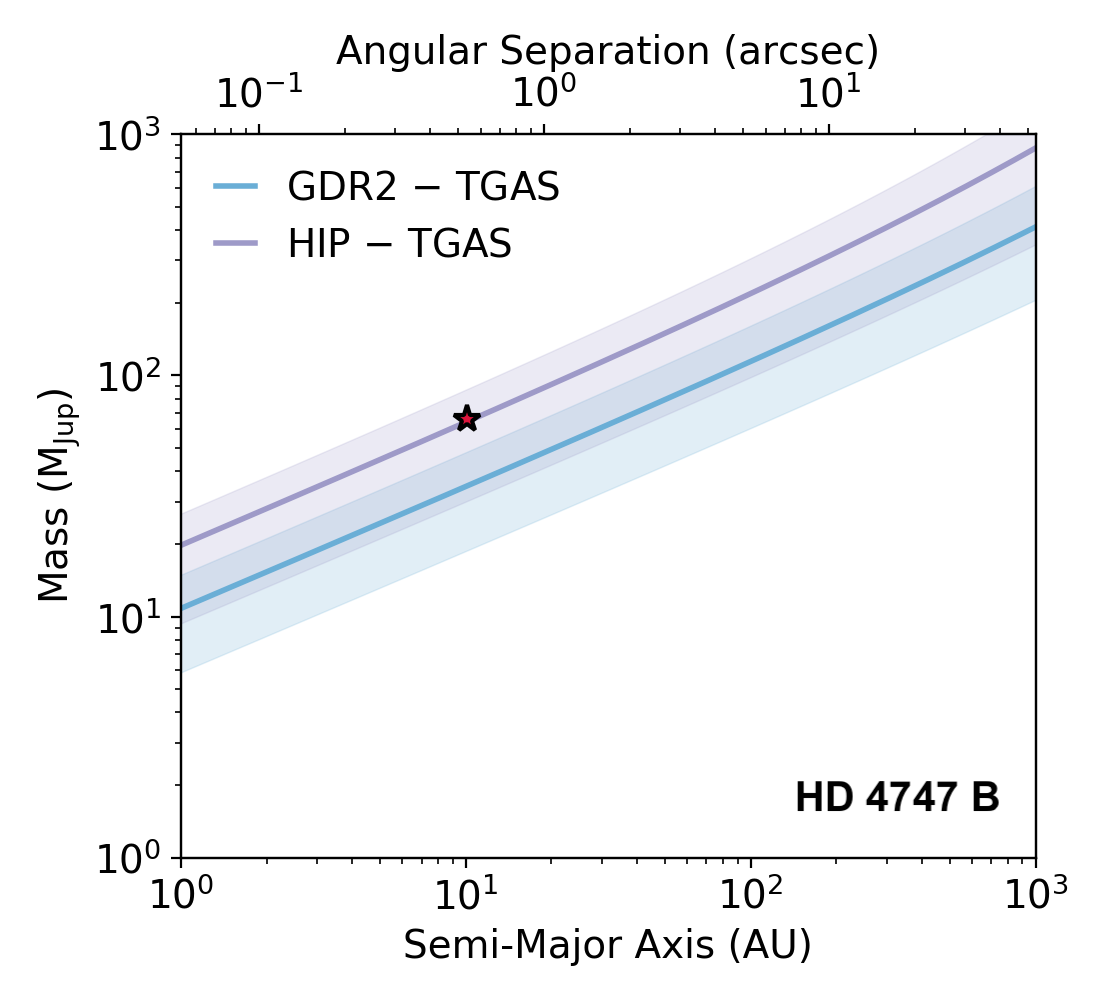}
    \end{minipage}
    \\
    \begin{minipage}[t]{0.445\textwidth}
        \includegraphics[width=\textwidth]{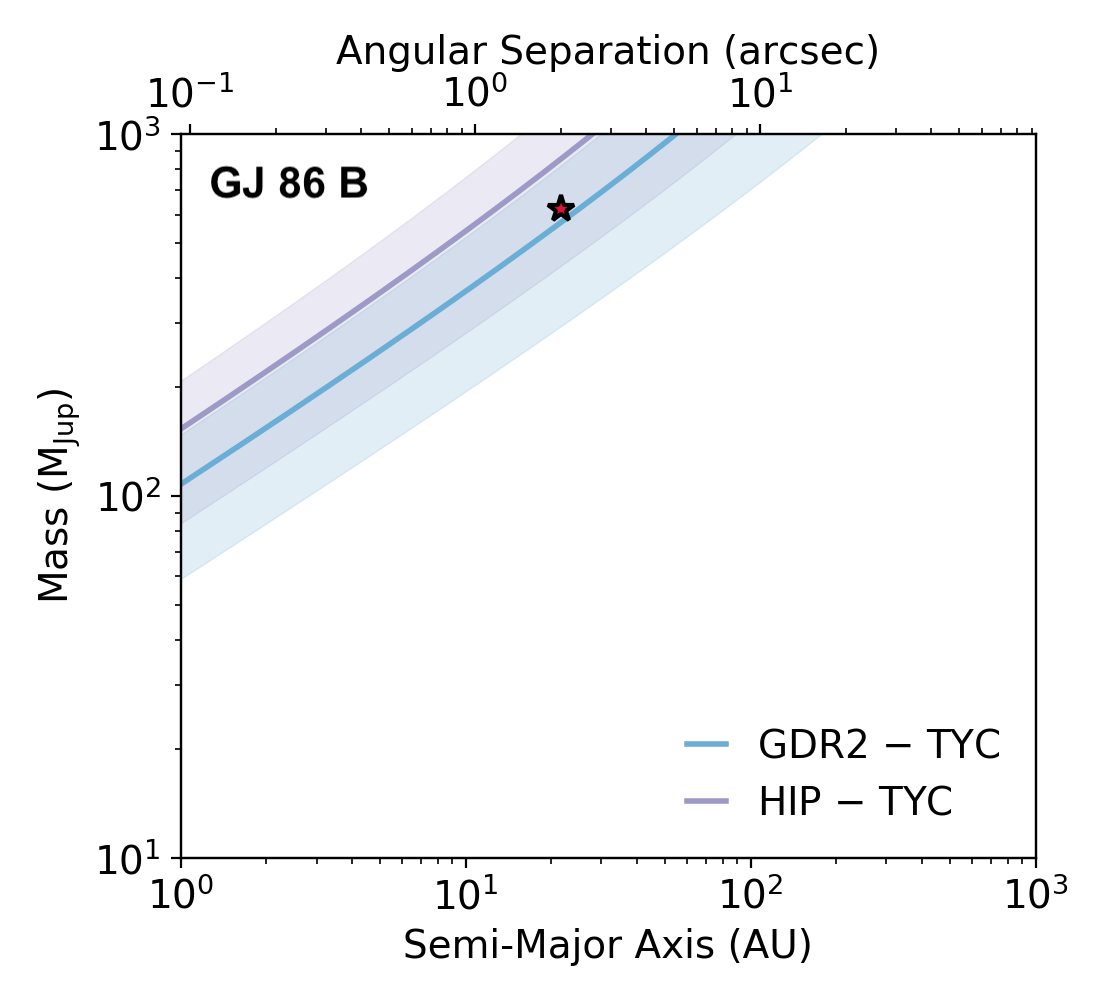}
    \end{minipage} \hspace{0.8cm} %\hfill
    \begin{minipage}[t]{0.445\textwidth}
        \includegraphics[width=\textwidth]{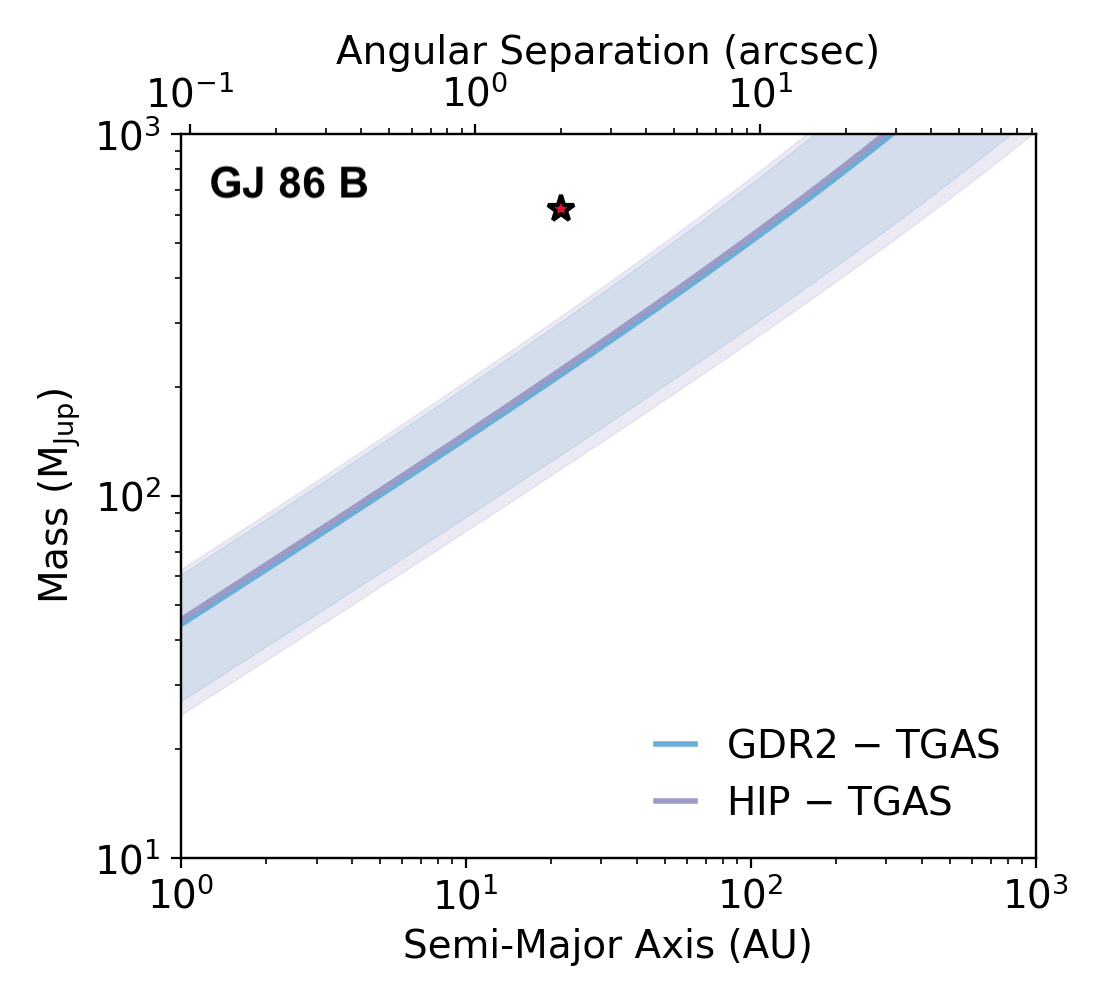}
    \end{minipage}
    \\
    \begin{minipage}[t]{0.445\textwidth}
        \includegraphics[width=\textwidth]{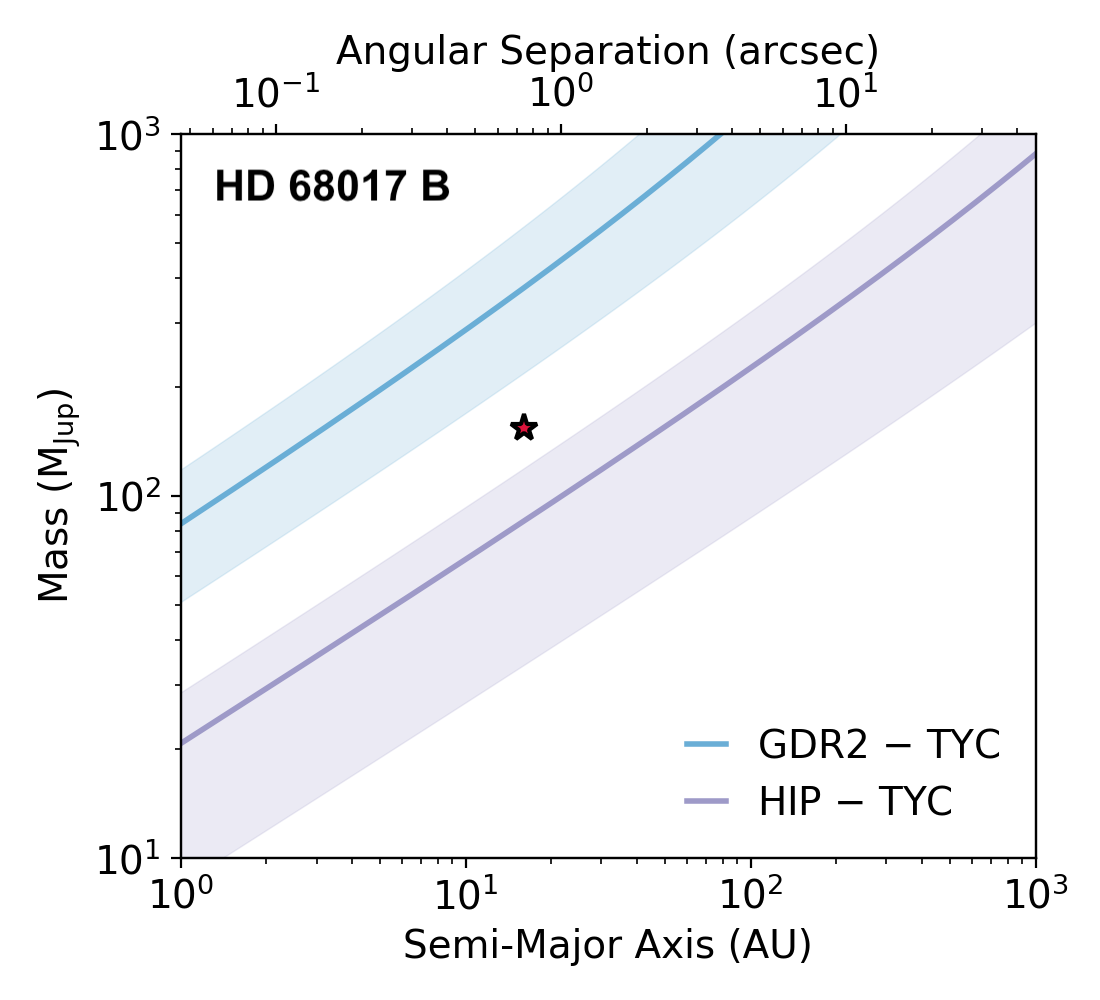}
    \end{minipage} \hspace{0.495\textwidth} %\hfill
    \caption{Output of COPAINS for the five targets studied in \citet{Brandt2018}, using \textit{Gaia} DR2 (blue) and \textit{Hipparcos} (purple) as short-term proper motion measurements, and Tycho-2 (left panels) and TGAS (right panels) for long-term proper motions. The solid lines and shaded areas correspond to the median and 1-$\sigma$ intervals, assuming a flat distribution in eccentricity. In general, the obtained results for the dynamical constrained placed with our code were found to be in good agreement with the positions of the secondary companions responsible for the observed astrometric trends (red stars). The error bars on the semi-major axes and masses of the companions are plotted but are in most cases smaller than the plot symbol. The solutions marked by hatched areas correspond to $\Delta\mu$ values with low significances (see Table \ref{t:targets_table}), which would not normally be included in such an analysis.}
    \label{f:DMU_all}
\end{figure*}
\begin{figure*}
\addtocounter{figure}{-1}
    \centering
    \begin{minipage}[t]{0.445\textwidth}
        \includegraphics[width=\textwidth]{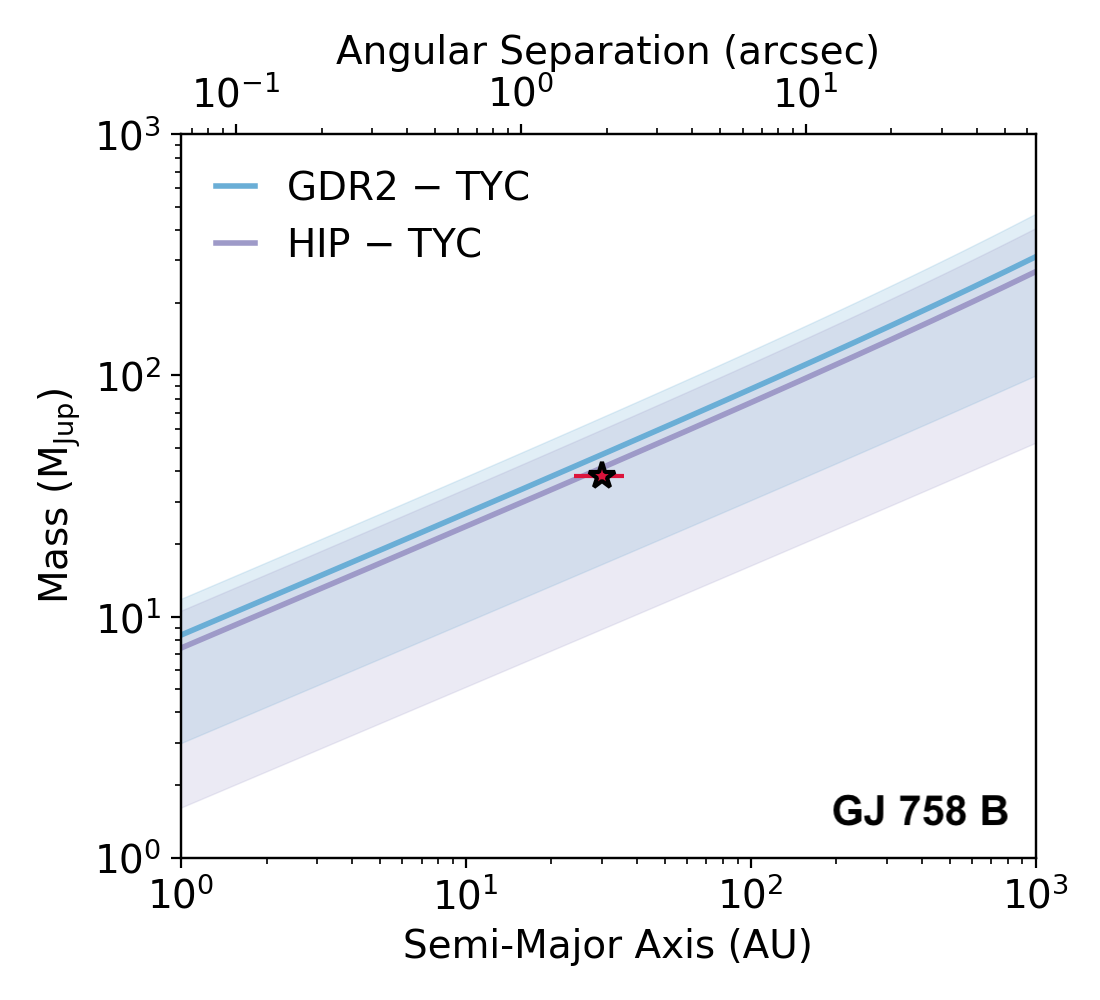}
    \end{minipage} \hspace{0.8cm} %\hfill
    \begin{minipage}[t]{0.445\textwidth}
        \includegraphics[width=\textwidth]{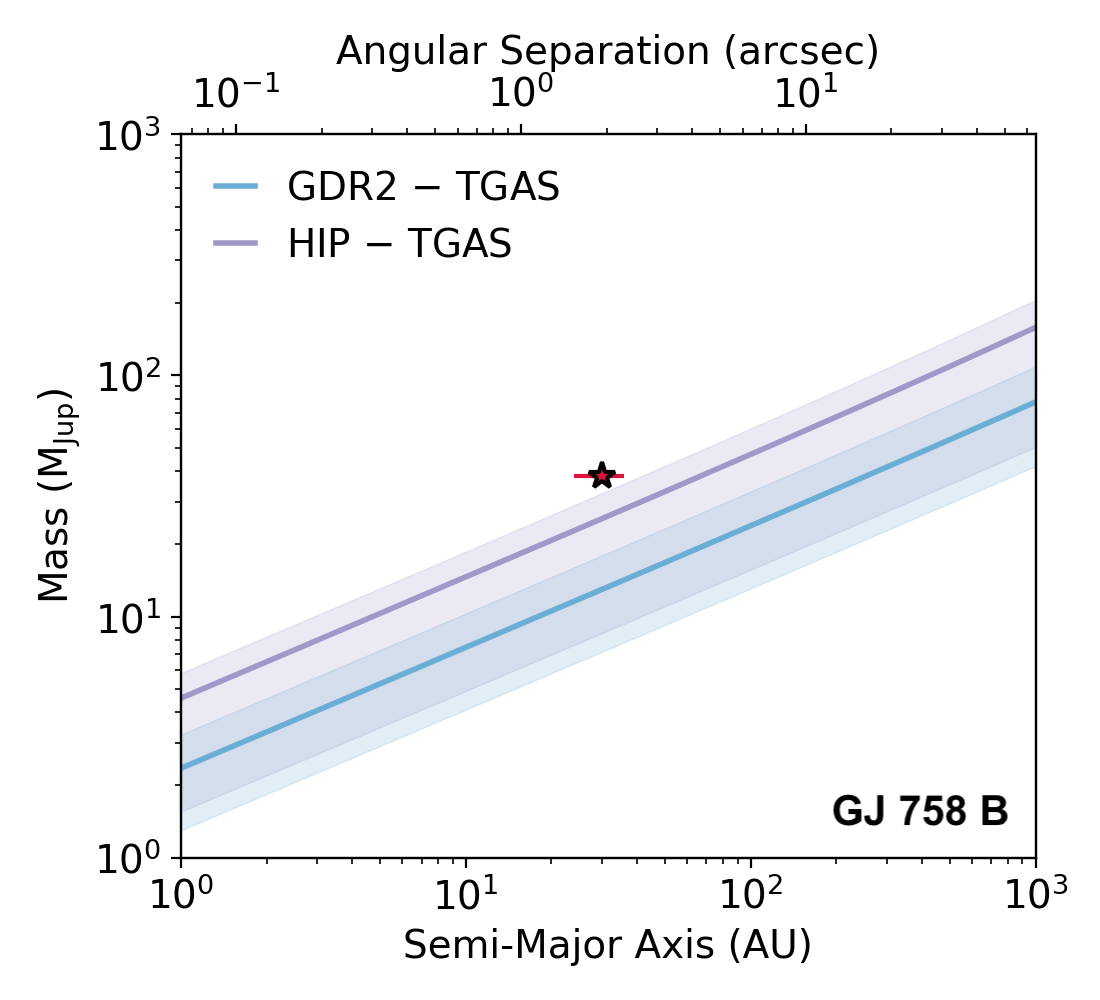}
    \end{minipage}
    \\ %\vspace{0.5cm}
    \begin{minipage}[t]{0.445\textwidth}
        \includegraphics[width=\textwidth]{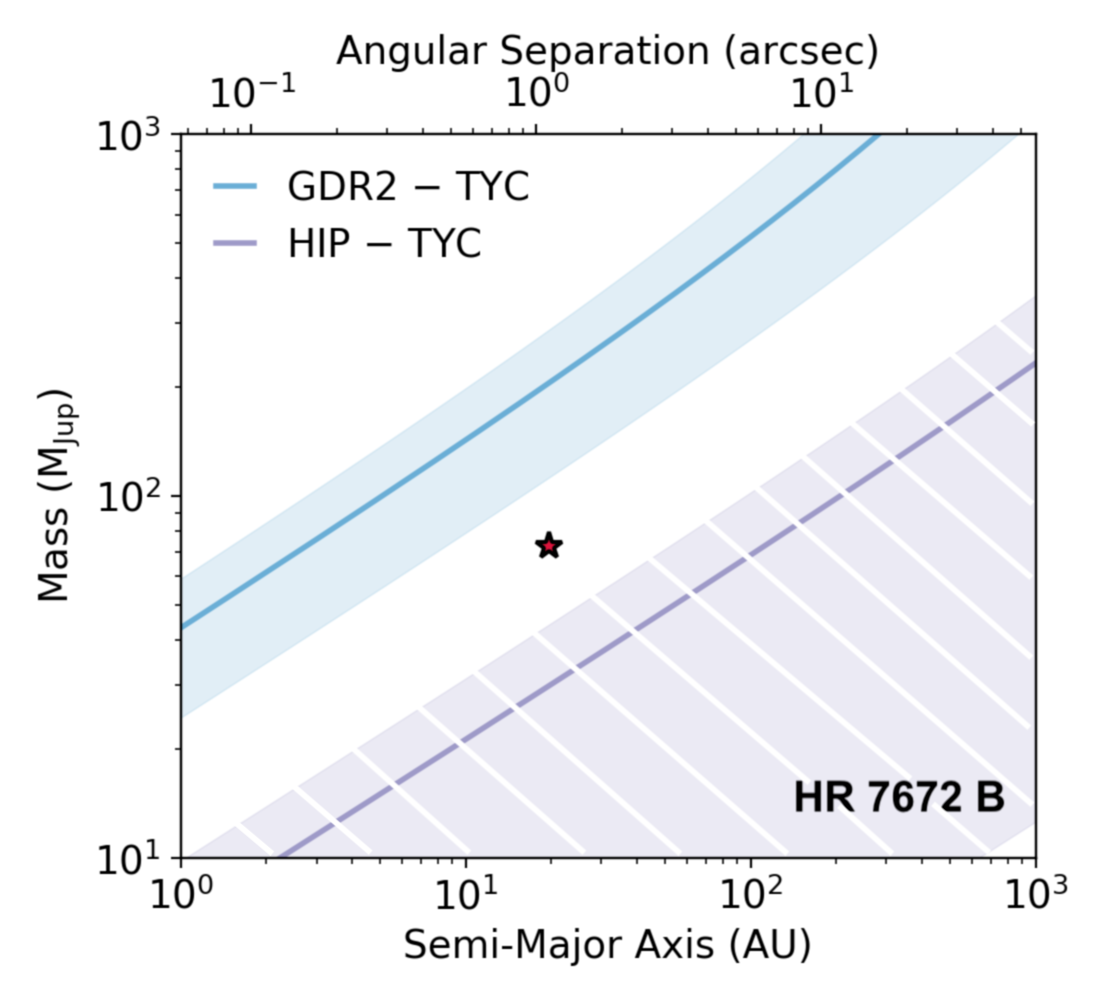}
    \end{minipage} \hspace{0.8cm} %\hfill
    \begin{minipage}[t]{0.445\textwidth}
        \includegraphics[width=\textwidth]{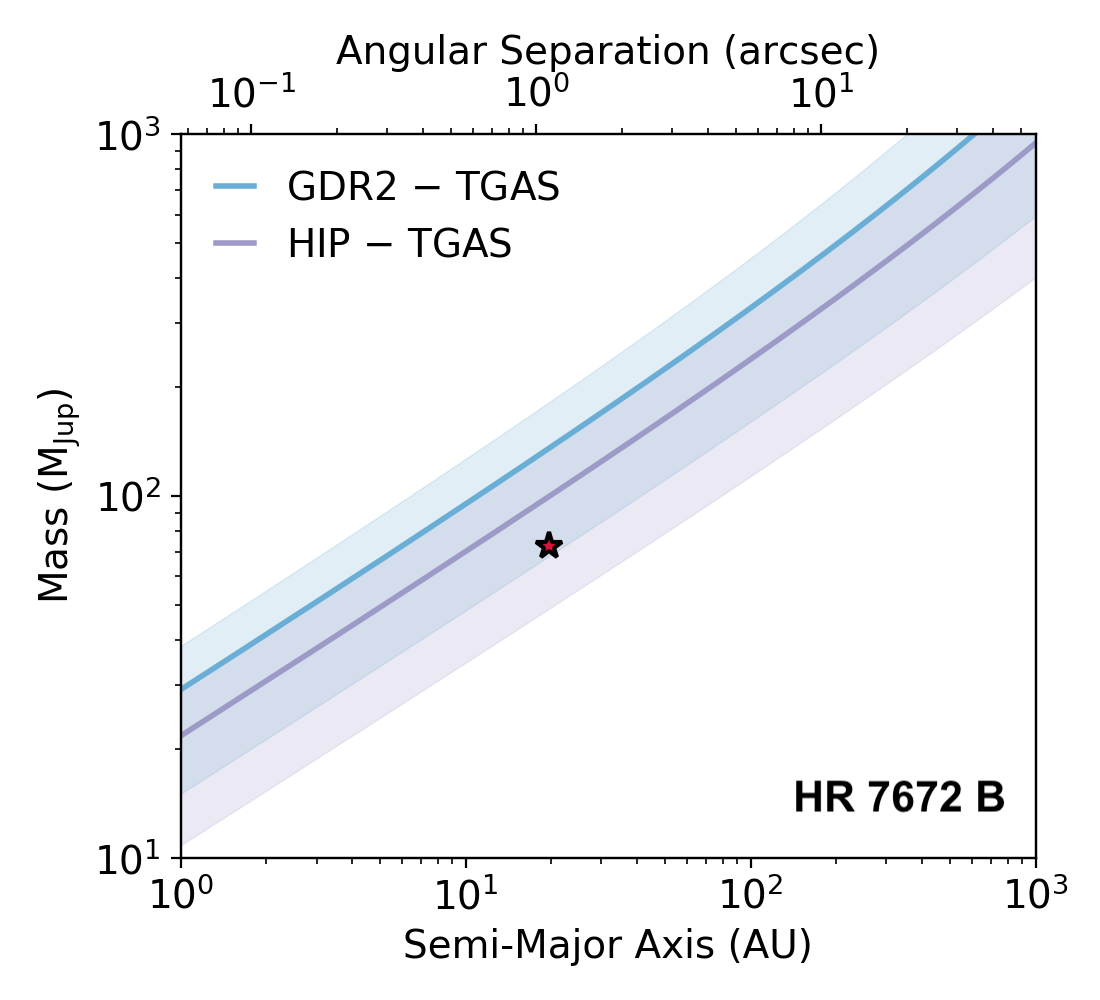}
    \end{minipage}
    \caption{(Continued.)}
\end{figure*}
}

From Equation \ref{eq:dmu_measured}, we estimated $\Delta\mu$ values for each target using the TGAS and Tycho-2 catalogues as long-term proper motion measurements, and adopting the \textit{Gaia} DR2 and \textit{Hipparcos} data as instantaneous velocities. For \textit{Hipparcos}, we consider the new reduction of the raw data \citep{vanLeeuwen2007}, as those data were used in the TGAS solutions. All systems have at least one significant measurement of a change in proper motion based on these catalogues. Table \ref{t:targets_table} summarises the physical properties and astrometric information for each system. This sample provides a range of secondary companions, with masses extending from the substellar regime to low-mass stars, in addition to the white dwarf companion to GJ 86. With orbital periods varying between 30 and 180 yr for these systems, both \textit{Gaia} DR2 and \textit{Hipparcos} are expected to provide good approximations to the instantaneous velocities of the primaries. TGAS and Tycho-2 proper motions, on the other hand, are likely to be somewhat discrepant from the proper motions of the systems' barycentres given the long orbital periods of some of these targets (see Section \ref{limitations}).

Using the information from Table \ref{t:targets_table}, we ran our COPAINS tool on each system and for each observed $\Delta\mu$ measurement. We adopted uniform distributions in eccentricity based on the spread of eccentricity values observed among the five targets. The output of the code for each combination of long and short-term astrometric data is presented in Figure \ref{f:DMU_all} for all systems. The solid lines show the median set of solutions and the shaded areas correspond to 68\% confidence levels. The red stars indicate the positions of the companions based on the results from \citet{Brandt2018}, allowing us to compare the predictions from COPAINS to reliable semi-major axes and dynamical mass estimates for each system.

Most trends were found to be consistent with the positions of the secondary companions at the 2-$\sigma$ level, with about half agreeing within 1$\sigma$ (shaded areas in the plots). This suggests that the uncertainties introduced by the time baselines of the proper motions used are in most cases not highly significant, in agreement with our analysis in Section \ref{catalogue_length}. Tycho-2 was typically found to provide a better estimate of the centre-of-mass motions, as a result of its longer temporal coverage (about a century) compared to the $\sim$25-yr baseline of TGAS. Since the magnitude of the observed $\Delta\mu$ varies with orbital phase, the observed offsets between the usage of \textit{Gaia} DR2 and \textit{Hipparcos} as short-term measurements could (partly) reflect different phases at each individual epoch.

As expected, larger uncertainties in the measured $\Delta\mu$ resulted in wider distributions for a given confidence interval. Low-significance $\Delta\mu$ values (marked with an asterisk in Table \ref{t:targets_table}) produced a broad lower tail, as shown by the hatched areas in Figure \ref{f:DMU_all}. The various combinations of catalogues considered were also generally found to be consistent with each other for most systems, with the exception of these low-significance trends.

Beside larger uncertainties on comparable measurements, these can arise from cases where the short-term and long-term proper motion vectors are close to alignment. As noted in \citet{Makarov2005}, this can occur in eccentric astrometric binaries, where the binary components spend most of their time at large separations, with low orbital velocities.
While we considered these values here for the completeness of our analysis, such low-significance measurements would not normally be selected for an inspection with COPAINS. Only the narrower trends, providing strong constraints on the region of the parameter space in which the secondary companions may be located, would be used for a direct imaging target selection. No additional offset was found for the targets with the largest orbital inclinations (e.g. GJ 86, HR 7672) relative to very low-inclination systems (e.g. HD 68017), confirming that inclination will statistically rarely have a large impact on the results from our code, as we showed in Section \ref{inclination}. 

We conclude that our method allows for an efficient estimate of the probable location of a hidden companion based on astrometric displacements of its host star. Our tool was found to yield a good indication of the region of the parameter space of interest, despite the uncertainties introduced by assumptions made in our approach, and the disparities originating from the use of different astrometric catalogues.

\section{Conclusions and future prospects}
\label{conclusion}

We presented in this paper a code aimed at identifying new directly-imagable companions. Our COPAINS tool (Code for Orbital Parametrisation of Astrometrically Inferred New Systems) exploits the synergy between direct imaging and astrometry with the aim to select the best targets for imaging campaigns searching for low-mass companions. Our approach is based on changes in stellar proper motions ($\Delta\mu)$ induced by the gravitational influence of a hidden companion. We have shown that stars with significant proper motion accelerations have a much higher binary fraction than stars that do not exhibit any astrometric excursion, confirming the efficiency of such a selection procedure for companion searches. We also found that the precision of \textit{Gaia} astrometric measurements from the DR1 and DR2 catalogues is sufficient to uncover about half of currently-known directly-imaged substellar companions to single stars based on this selection method. This number is expected to improve significantly with the more complete and more precise future \textit{Gaia} Data Releases, which will make our method sensitive to astrometric trends caused by very low-mass planetary companions.

From measured astrometric offsets between long and short-term proper motion measurements, our COPAINS tool allows for the computation of the secondary mass and separation pairs compatible with the observed trend, marginalised over all possible orbital phases and eccentricities. The resulting solutions are based entirely of dynamical arguments, although a dependence on the adopted (usually model-derived) stellar mass remains in the obtained secondary masses. Testing our method on systems with well-characterised orbits, we found that our approach provides a good indication of the location of invisible companions responsible for the positional offsets of their host stars. By comparing the output of the code to the typical performance of imaging facilities or existing sets of observations, COPAINS offers a robust, informed selection process for ideal targets to observe in direct imaging surveys. 

The results obtained here are extremely encouraging regarding the discovery of new direct imaging companions in the coming years, especially with the advent of the next generation of telescopes. Samples gathered with such an informed selection procedure are likely to considerably increase the very low detection rates from current imaging campaigns \citep{Bowler2016}. The anticipated \textit{James Webb Space Telescope} (\textit{JWST}) will allow for unparalleled probes of planetary atmospheres, but will primarily serve as a characterisation mission. It is thus vital to start identifying new benchmark systems to follow-up with \textit{JWST} in anticipation of its commissioning.
In addition, null detections with current-generation facilities (e.g. VLT/SPHERE, Gemini/GPI) for surveys selected with COPAINS would also allow us to rule out the presently-accessible part of the parameter space for the observed targets. With evidence of a perturbing body and no detection at currently-probed masses and separations, such results would yield excellent samples for ground-based direct imaging surveys with the upcoming Extremely Large Telescopes (ELTs), significantly reducing the risks of non-detections.

\section*{Acknowledgements}
The authors thank the anonymous referee for a report that helped improve the quality of the paper. The authors would also like to thank M. Janson for helpful conversations and comments regarding the manuscript, as well as O. Absil and C. Delacroix for their help with the METIS contrast curve. This study was supported by a Research Incentive Grant from the Carnegie Trust for the Universities of Scotland.
KM acknowledges funding by the Science and Technology Foundation of Portugal (FCT), grants No. IF/00194/2015 and PTDC/FIS-AST/28731/2017.
This work has made use of data from the European Space Agency (ESA) mission
{\it Gaia} (\url{https://www.cosmos.esa.int/gaia}), processed by the {\it Gaia} Data Processing and Analysis Consortium (DPAC, \url{https://www.cosmos.esa.int/web/gaia/dpac/consortium}).

\bibliographystyle{mnras}
\bibliography{biblio}

%%%%%%%%%%%%%%%%%%%%%%%%%%%%%%%%%%%%%%%%%%%%%%%%%%

\bsp	% typesetting comment
\label{lastpage}

\end{document}